\documentclass[]{aa}

\usepackage{bm}
\usepackage{amsmath}
\usepackage{xcolor}
\usepackage{graphicx}
\usepackage{epstopdf}
\usepackage{hyperref}
\usepackage[all]{hypcap} 
\usepackage{dblfloatfix}    

\begin{document}
	
	\title{Two-dimensional simulations of coronal rain dynamics}
	\subtitle{I. Model with vertical magnetic field and an unbounded atmosphere}

	\author{D. Martínez-Gómez\inst{\ref{inst1}}
		\and R. Oliver\inst{\ref{inst2},\ref{inst3}}
		\and E. Khomenko\inst{\ref{inst1},\ref{inst4}}
		\and M. Collados\inst{\ref{inst1},\ref{inst4}}}
	
	\institute{Instituto de Astrofísica de Canarias, 38205 La Laguna, Tenerife, Spain; \email{dmartinez@iac.es}\label{inst1}
		\and Departament de Física, Universitat de les Illes Balears, E-07122 Palma de Mallorca, Spain\label{inst2}
		\and Institute of Applied Computing \& Community Code (IAC3), UIB, Spain\label{inst3}
		\and Departamento de Astrofísica, Universidad de La Laguna, 38205 La Laguna, Tenerife, Spain\label{inst4}
	}

	\abstract{Coronal rain often arises as the final product of evaporation and condensation cycles that occur in active regions. Observations have shown that the condensed plasma falls with an acceleration smaller than that of free fall.}{We aim to improve the understanding of the physical mechanisms behind the slower than free-fall motion and the two-stage evolution (an initial phase of acceleration followed by an almost constant velocity phase) detected in coronal rain events.}{Using the \textsc{Mancha3D} code, we solve the 2D ideal MHD equations. We represent the solar corona as an isothermal vertically stratified atmosphere with a uniform vertical magnetic field and the plasma condensation as a density enhancement described by a 2D Gaussian profile. We analyse the temporal evolution of the descending plasma and study its dependence on parameters such as density and magnetic field strength.}{We confirm previous findings that the pressure gradient is the main force that opposes the action of gravity and slows down the blob descent and that larger densities require larger pressure gradients to reach the constant speed phase. We find that the shape of a condensation with a horizontal variation of density is distorted as it falls, due to the denser parts of the blob falling faster than the lighter ones. This is explained by the fact that the duration of the initial acceleration phase, and therefore the maximum falling speed attained by the plasma, increases with the ratio of blob to coronal density. We also find that the magnetic field plays a fundamental role in the evolution of the descending condensations. A strong enough magnetic field (greater than 10 G in our simulations) forces each plasma element to follow the path given by a particular field line, which allows to describe the evolution of each vertical slice of the blob in terms of 1D dynamics, without influence of the adjacent slices. In addition, under the typical conditions of the coronal rain events, the magnetic field prevents the development of the Kelvin-Helmholtz instability.}{}
	
	
	\maketitle
	
	\titlerunning{2D simulations of coronal rain}
	\authorrunning{Martínez-Gómez et al.}

\section{Introduction}
    Coronal rain is a phenomenon that occurs in coronal loops of active regions. It consists in the condensation of the hot and dim coronal plasma into cold elongated blobs that then fall towards the solar surface following the leg of the loop. It was already more than fifty years ago when the first observational evidences of this phenomenon were presented by \citet{1962AuJPh..15..327B}. This and subsequent works such as \citet{1969SoPh....8...29B,1978ApJ...223.1046F} or \citet{1979SoPh...62..331E} reported falling speeds ranging from $40$ to $60 \ \rm{km \ s^{-1}}$ but did not study the kinematics of the condensations. Later, \citet{1984ApJ...283..392L,1987sman.work..147X} and \citet{1992SoPh..139...81H} performed a tracking of the positions and speeds of the blobs as functions of height and determined that they followed a slower than free-fall motion. 
    
    More recent investigations measured velocities of around $30-40 \ \rm{km \ s^{-1}}$ near the top of the loops, but values of up to $150 \ \rm{km \ s^{-1}}$ further down the loop leg \citep{1996SoPh..166...89W,2004A&A...415.1141D,2005A&A...436.1067M,2010ApJ...716..154A,2012ApJ...745..152A}. These observations, in addition to the works of \citet{2001SoPh..198..325S,2009RAA.....9.1368Z} and \citet{2012SoPh..280..457A}, revealed that two stages can be distinguished in the motion of the condensations. The plasma starts falling with an acceleration that is smaller than that of gravity and decreases until an almost constant speed phase is reached. Therefore, there must be a physical mechanism that opposes the action of gravity and considerably slows down the descent of the cold blobs. This two-stage feature is not unique for coronal rain events but a similar behaviour has also being detected in the descent of plasma after a prominence eruption \citep{2014NewA...26...23X}. 
    
    The evidence that informs about the low temperature of these blobs comes from the fact that they are typically observed in lines associated with cold plasma such as $\rm{Ly\alpha}$, $\rm{H}\alpha$ or $\rm{He} \ \textsc{ii} \ 304$ \AA{} but do not appear in hot coronal lines like $171$ \AA{} \citep{2004A&A...415.1141D,2005A&A...436.1067M}. At these temperatures, usually found in the chromosphere and the transition region, the plasma is not expected to be fully ionised but to also contain neutral species. The presence of this neutral component has been demonstrated by the observations of \citet{2010ApJ...714..618C} and \citet{2014SoPh..289.4117A}. Regarding how the coronal plasma cools to such low temperatures and the condensation forms, several explanations have been developed. For instance, \citet{2011ApJ...736..121A} proposed that a large density enhancement can appear in a transversally oscillating coronal loop due to the ponderomotive force associated with large amplitude oscillations \citep{2004ApJ...610..523T}. Some alternative mechanisms involve the presence of magnetic null points and explain the creation of the dense blob either by a pressure pulse that excites an entropy mode \citep{2011A&A...533A..18M} or by interchange reconnection \citep{2019ApJ...874L..33M}.
    
    However, most models that try to replicate the formation of coronal rain take the catastrophic cooling due to a thermal instability \citep{1953ApJ...117..431P,1965ApJ...142..531F} as the main ingredient for their recipe. In these scenarios, heating concentrated at the loops base produces the evaporation of the plasma from the solar surface that then starts filling the magnetic structures. The heating can be impulsive or steady. In the former case, an intense burst of heating, like that caused by a flare event, stops abruptly and, due to the radiative losses, the plasma cools down to chromospheric temperatures \citep{1980ApJ...241..385A}. This mechanism explains coronal rain events in post-flare loops \citep{2015ApJ...803...85L,2016ApJ...833..184S}, in which blobs form and fall only once. But it does not explain how an active region can undergo several cycles of coronal rain. Observations by \citet{2012ApJ...745..152A,2015ApJ...806...81A,2015ApJ...807..158F,2017ApJ...835..272F} and \citet{2018ApJ...853..176A} have shown that the lifetime of the magnetic field structure is larger than the time taken for a condensation to form and fall, and that numerous blobs can consecutively descend along the same loop path. Such behaviour is better understood if the heating is assumed to be steady (or with a high frequency) and to decrease with height. In this case, the hot evaporated plasma starts to accumulate at the apex of the loop, where the heat input is not enough to maintain the energy balance with the radiative losses. This situation is referred to as a state of thermal non-equilibrium \citep[TNE;][]{1999ApJ...512..985A}. Radiative losses increase with decreasing temperatures and increasing densities, leading to a catastrophic cooling and the fast condensation of the plasma. Then, since the condensation is not in mechanical equilibrium, it starts falling down under the action of gravity. The validity of this TNE scenario, which can also be applied to explain the origin of solar prominences \citep{1991ApJ...378..372A}, has been demonstrated by both hydrodynamic \citep{1982A&A...108L...1K,1983A&A...123..216M,2001ApJ...553L..85K,2004A&A...424..289M} and magnetohydrodynamic \citep{2013ApJ...771L..29F,2015ApJ...807..142F,2015AdSpR..56.2738M} simulations.
    
    Regarding the reasons behind the smaller than free-fall motion, \citet{2001SoPh..198..325S} suggested that the speed of the condensations is not determined by gravity but by the evolution of the internal pressure of the loops. This suggestion was later supported by the works of \citet{2010ApJ...716..154A}, who reached the conclusion that the structure and dynamics of the blobs are more sensitive to variations in pressure than to gravity, and \citet{2014ApJ...784...21O}, who reproduced the two-stage evolution with a 1D numerical model and demonstrated the role of the pressure gradient in slowing down the plasma fall.
    
    The previous paragraphs have hinted several reasons why the study of coronal rain is a very relevant topic in the field of solar physics. It can improve the understanding of the processes involved in coronal heating and thermal instabilities. Then, coronal rain events can be used as magnetic field tracers and help to better describe the multistranded structure of the coronal magnetic field \citep{2014ApJ...792...93H,2014ApJ...797...36S}. Furthermore, the low temperatures of the descending blobs allow to explore the interaction between the ionised and the neutral components of partially ionized plasmas \citep{2016ApJ...818..128O}. 
    
    The present work is the first of a series in which we aim at obtaining a better comprehension of the dynamics of coronal rain. This is a quite complex task, as evidenced by the amount of past research devoted to the study of this phenomenon. So, instead of trying to reproduce every phase of the coronal rain cycle, here we follow the same approach as in \citet{2014ApJ...784...21O} of focusing only on its final stages i.e., on the descent of the condensation towards the solar surface. This will allow us to explore in great detail the forces that determine the evolution of the falling condensations. In this series, we improve the numerical 1D model used in \citet{2014ApJ...784...21O,2016ApJ...818..128O} by taking into account the 2D evolution of the falling blobs and including the effects of the presence of the magnetic field. In the first installment of the series, we  start by assuming that the plasma is fully ionised and use a single-fluid numerical code to compute its temporal evolution. We use the simple model of an unbounded isothermal atmosphere with a vertical magnetic field and investigate the dependence of the blob motion on parameters such as its density and the magnetic field strength. We also analyse the conditions that could lead to development of the Kelvin-Helmholtz instability \citep{1961hhs..book.....C} during the plasma descent. In the forthcoming works, we will add new features to this model, such as more complex configurations of the background magnetic field or the inclusion of a large density enhancement at the lower boundary to represent the chromosphere. In addition, we will use a two-fluid code to take into account the fact that the plasma in the condensations is partially ionised.
    
	This paper is organised as follows. In Section \ref{sec:model} we present the equations that govern the motion of the plasma and the numerical setup we use in our simulations. Section \ref{sec:results} contains the analysis of the simulations we have performed, which includes the study of the kinematics and dynamics of the falling blobs, in addition to the research of the possible development of instabilities. Finally, the conclusions of this investigation are summarised in Section \ref{sec:concl}.
	
\section{Model} \label{sec:model}
\subsection{Governing equations}
    For the present research, we have assumed that the plasma is fully ionised. To analyse its dynamics, we have considered the single-fluid approximation and have neglected non-ideal effects. Therefore, the temporal evolution of the density $(\rho)$, velocity $(\bm{V})$, energy $(E)$ and magnetic field $(\bm{B})$ is described by the set of ideal magnetohydrodynamics (MHD) equations:
	\begin{equation} \label{eq:cont}
		\frac{\partial \rho}{\partial t}+\nabla \cdot \left(\rho \bm{V}\right)=0,
	\end{equation} 
	\begin{equation} \label{eq:momen}
		\frac{\partial \left(\rho \bm{V}\right)}{\partial t}+\nabla \cdot \left[\rho \bm{V}\bm{V}+\left(P+\frac{\bm{B}^{2}}{2\mu_{0}}\right) \mathbb{I}-\frac{\bm{B}\bm{B}}{\mu_{0}}\right]=\rho \bm{g},
	\end{equation} 
	\begin{equation} \label{eq:energy}
		\frac{\partial E}{\partial t}+\nabla \cdot \left[\left(E+P+\frac{\bm{B}^{2}}{2\mu_{0}}\right)\bm{V}-\frac{\bm{B}}{\mu_{0}}\left(\bm{B}\cdot \bm{V}\right)\right]=\rho \bm{V}\cdot \bm{g},
	\end{equation} 
	\begin{equation} \label{eq:induction}
		\frac{\partial \bm{B}}{\partial t}=\nabla \times \left(\bm{V} \times \bm{B}\right),
	\end{equation} 
	where $P$ is the pressure, $\mu_{0}$ is the magnetic permeability, $\mathbb{I}$ is the identity tensor  and $\bm{g}$ is the gravitational acceleration. The total energy is defined as
	\begin{equation} \label{eq:ene_def}
		E=\frac{1}{2}\rho \bm{V}^{2}+\frac{P}{\gamma-1}+\frac{\bm{B}^{2}}{2\mu_{0}},
	\end{equation} 
	where $\gamma$ is the adiabatic index.
	
	In addition, to close the system of equations we use the perfect gas equation, which is given by
	\begin{equation} \label{eq:state}
		P =\frac{k_{\rm{B}}}{\widetilde{\mu}m_{\rm{p}}}\rho T,
	\end{equation} 
	where $m_{\rm{p}}$ is the proton mass, $k_{\rm{B}}$ is the Boltzmann constant, $T$ is the temperature, and $\widetilde{\mu}$ is the mean atomic weight. In this work, we assume that the plasma is only composed of hydrogen, which corresponds to a value of $\widetilde{\mu} = 0.5$.
	
\subsection{Numerical setup}
	All the simulations analysed in the present work have been carried out with the 
	\textsc{Mancha3D} code \citep{2006ApJ...653..739K,2010ApJ...719..357F,2018A&A...615A..67G}. This numerical code solves the non-ideal non-linear equations of single-fluid MHD in three dimensions. The non-ideal terms include the ambipolar diffusion due to the presence of neutrals in the plasma, the Hall effect, the battery effect and radiative losses. However, for the present investigation we have neglected the dependence on the $y$-direction, so our simulations are two-dimensional, and we have not taken into account those non-ideal effects but have retained only the terms that correspond to the ideal MHD description.
	
    The \textsc{Mancha3D} code treats each variable in the system of Eqs. (\ref{eq:cont})--(\ref{eq:induction}) as a combination of a static equilibrium value plus a perturbation (that can be non-linear), and computes the temporal evolution of the perturbations. Therefore, for instance, the variables density and pressure can be written as
    \begin{equation}
    	\rho = \rho_{\rm{eq}} + \rho_{1}, \hspace{0.5cm} P = P_{\rm{eq}} + P_{1},
    \end{equation}
    where $\rho_{\rm{eq}} = \rho_{\rm{eq}}(x,z)$ and $P_{\rm{eq}} = P_{\rm{eq}}(x,z)$ are the equilibrium values, and $\rho_{1} = \rho_{1}(x, z, t)$ and $P_{1} = P_{1}(x, z, t)$ are the perturbations, respectively.
    
    The code uses a centered $6^{th}$ order scheme for spatial integration and the temporal evolution is computed by means of an explicit 4-step Runge-Kutta method. To prevent the exponential growth of high-frequency noise related to the small scales of the numerical grid, artificial diffusion terms are added to the right-hand side of Eqs. (\ref{eq:cont})--(\ref{eq:induction}). This artificial diffusion is formed by a shock resolving term and a hyperdiffusivity part, as described in \citet{2005A&A...429..335V}.
   
\subsubsection{Equilibrium state}
	For the equilibrium state in our simulations, we consider the solar corona as an isothermal vertically stratified atmosphere, with a temperature given by $T_{0}$ and a uniform vertical magnetic field given by $\bm{B_{0}}=(0, 0, B_{0})$, where the $z$-axis points upward. Mathematically, this equilibrium is described as follows. From Eq. (\ref{eq:momen}), if we ignore the temporal variations, assume $\bm{V} = \vec{0}$ and consider the variations along the $z$-direction, we get that
    \begin{equation}
    	0 = -\frac{\partial P_{\rm{eq}}}{\partial z}-\rho_{\rm{eq}} g,
    \end{equation}
    which, using Eq. (\ref{eq:state}), transforms into
    
    \begin{equation}
    	\frac{\partial \rho_{\rm{eq}}}{\partial z}=-\frac{\widetilde{\mu}m_{\rm{p}}g}{k_{\rm{B}}T_{0}}\rho_{\rm{eq}}.
    \end{equation}
    
    Solving the previous equation, we get that the equilibrium state is described by
    \begin{equation} \label{eq:equil}
    	\rho_{\rm{eq}} = \rho_{0} e^{-z/H}, \hspace{0.5cm} P_{\rm{eq}} = P_{0} e^{-z/H},
    \end{equation}
    where $\rho_{0}$ and $P_{0}$ are the values of density and pressure at $z = 0$, respectively, and satisfy the relation $P_{0}= k_{\rm{B}}/(\widetilde{\mu}m_{p})\rho_{0}T_{0}$, and $H$ is the vertical scale height, given by
	\begin{equation}
		H=\frac{k_{\rm{B}}T_{0}}{\widetilde{\mu}m_{\rm{p}}g}.
	\end{equation} \label{eq:hscale}
	
	In all the simulations of the present study, the values of density and pressure at the base of the corona are $\rho_{0} = 5 \times 10^{-12} \ \rm{kg \ m^{-3}}$ and $P_{0} = 0.165 \ \rm{Pa}$. The chosen temperature is $T_{0} = 2 \times 10^{6} \ \rm{K}$, which, assuming the typical value of $g \simeq 274 \ \rm{m  \ s^{-2}}$ for the solar surface gravity, gives a vertical scale height $H \simeq 120 \ \rm{Mm}$.

\subsubsection{Initial and boundary conditions}
	At the initial step of our simulations, the blob is represented by a density enhancement, described by the following two-dimensional Gaussian profile:
	\begin{equation} \label{eq:blob}
		\rho_{\rm{b}}(x,z) = \rho_{\rm{b0}} \exp \left[-\frac{\left(x-x_{0}\right)^{2}+\left(z-z_{0}\right)^{2}}{\Delta^{2}}\right],
	\end{equation}
	where $\rho_{b0}$ is the density at the center of the blob, $x_{0}$ and $z_{0}$ represent its initial position and $\Delta$ is proportional to its width. This enhancement is inserted into the code through the variable corresponding to the perturbation of density, i.e., $\rho_{1}$. Therefore, the initial condition for the total density is given by $\rho (x, z, t = 0) = \rho_{\rm{eq}} + \rho_{1} (x, z, t=0) = \rho_{\rm{eq}} + \rho_{\rm{b}}(x, z)$.
	
	In the $x$-direction, we consider a physical domain of length $L_{x} = 10 \ \rm{Mm}$, that goes from $x_{\rm{left}} = -5 \ \rm{Mm}$ to $x_{\rm{right}} = 5 \ \rm{Mm}$. The height of the domain, $L_{z}$, and its bottom and top limits vary from one simulation to another, as we explain later. The simulation with the largest height uses a value of $L_{z} = 100 \ \rm{Mm}$. With a resolution of $10 \ \rm{km}$, this domain is represented by a mesh of $10^{3} \times 10^{4}$ points. We use periodic boundary conditions in the $x$-direction while in the vertical direction we use Perfectly Matched Layers \cite[PML;][]{1994JCoPh.114..185B,1996JCoPh.127..363B,2009ApJ...694..573P}. These layers prevent the reflection of perturbations at the boundaries, which may introduce unphysical behaviors in the system, and allow us to use a much smaller vertical-domain in comparison with that employed in \citet{2014ApJ...784...21O}.
	
\section{Results} \label{sec:results}
	According to the observations, once a dense blob is formed in the corona, it starts to fall towards the solar surface under the action of gravity. However, it follows a slower than free-fall motion. With our simulations we want to replicate this behaviour, to determine which mechanisms act against the gravity to slow down the fall and to study how the motion of the blob depends on parameters such as its density or the strength of the background magnetic field. For this reason, we have performed a series of simulations combining different values of the peak density of the blob, $\rho_{b0}$, and the magnetic field strength, $B_{0}$. Other parameters such as the width of the blob or its initial position, have been kept fixed throughout this series of simulations, with $\Delta = 500 \ \rm{km}$ and $x_{0} = 0$, $z_{0} = 50 \ \rm{Mm}$, respectively.
	
	\begin{table}
		\caption{Parameters of simulations}
		\label{table:sims}
		\centering
		\begin{tabular}{c c  c c c}
			\hline\hline
			Name & $\rho_{\rm{b0}} / \rho_{\rm{ref}}$ & $B_{0} \ (\rm{G})$ & $L_{z} (\rm{Mm})$ & Symbol \\
			\hline
			I-5 & 1 & 5 & 80 & $\textcolor{red}{\blacktriangle}$ \\
			I-10 & 1 & 10 & 80 & $\textcolor{green}{\blacktriangle}$ \\
			I-20 & 1 & 20 & 80 & $\textcolor{cyan}{\blacktriangle}$ \\
			V-0 & 5 & 0 & 100 &\\
			V-5 & 5 & 5 & 100 & $\textcolor{red}{\blacksquare}$ \\
			V-10 & 5 & 10 & 100 & $\textcolor{green}{\blacksquare}$ \\
			V-20 & 5 & 20 & 100 & $\textcolor{cyan}{\blacksquare}$ \\
			X-1 & 10 & 1 & 100 & \\
			X-5 & 10 & 5 & 100 & $\textcolor{red}{\blacklozenge}$ \\
			X-10 & 10 & 10 & 100 & $\textcolor{green}{\blacklozenge}$ \\
			X-20 & 10 & 20 & 100 & $\textcolor{cyan}{\blacklozenge}$ \\
			\hline
		\end{tabular}
		\tablefoot{$\rho_{\rm{ref}} = 10^{-10} \ \rm{kg \ m^{-3}}.$ The roman numeral in the abbreviated name of each simulation indicates the ratio $\rho_{\rm{b0}} / \rho_{\rm{ref}}$, while the arabic numeral indicates the magnetic field strength. The rightmost column shows the symbols used in Fig. \ref{fig:ratio} to represent the data obtained from each simulation.}
	\end{table}
	
	In Sects. \ref{sec:kin} and \ref{sec:dyn}, we analyse the results of a series of simulations in which we combine three values of the blob density ($\rho_{\rm{b0}} = 1, \ 5, \ 10 \times 10^{-10} \ \rm{kg \ m^{-3}}$) and three magnetic field strengths ($B_{0} = 5, \ 10, \ 20 \ \rm{G}$). In Section \ref{sec:inst}, we also include results from simulations with $B_{0} = 1 \ \rm{G}$ and without magnetic field. The set of parameters used in the different simulations are summarised in Table \ref{table:sims}. For the sake of simplicity, when discussing the results of a given simulation, instead of repeating the values of the chosen parameters, we will refer to it by the abbreviated name presented in Table \ref{table:sims}.
	
\subsection{2D blob kinematics} \label{sec:kin}
    \begin{figure*}
    	\centering
    	\includegraphics[width=17cm]{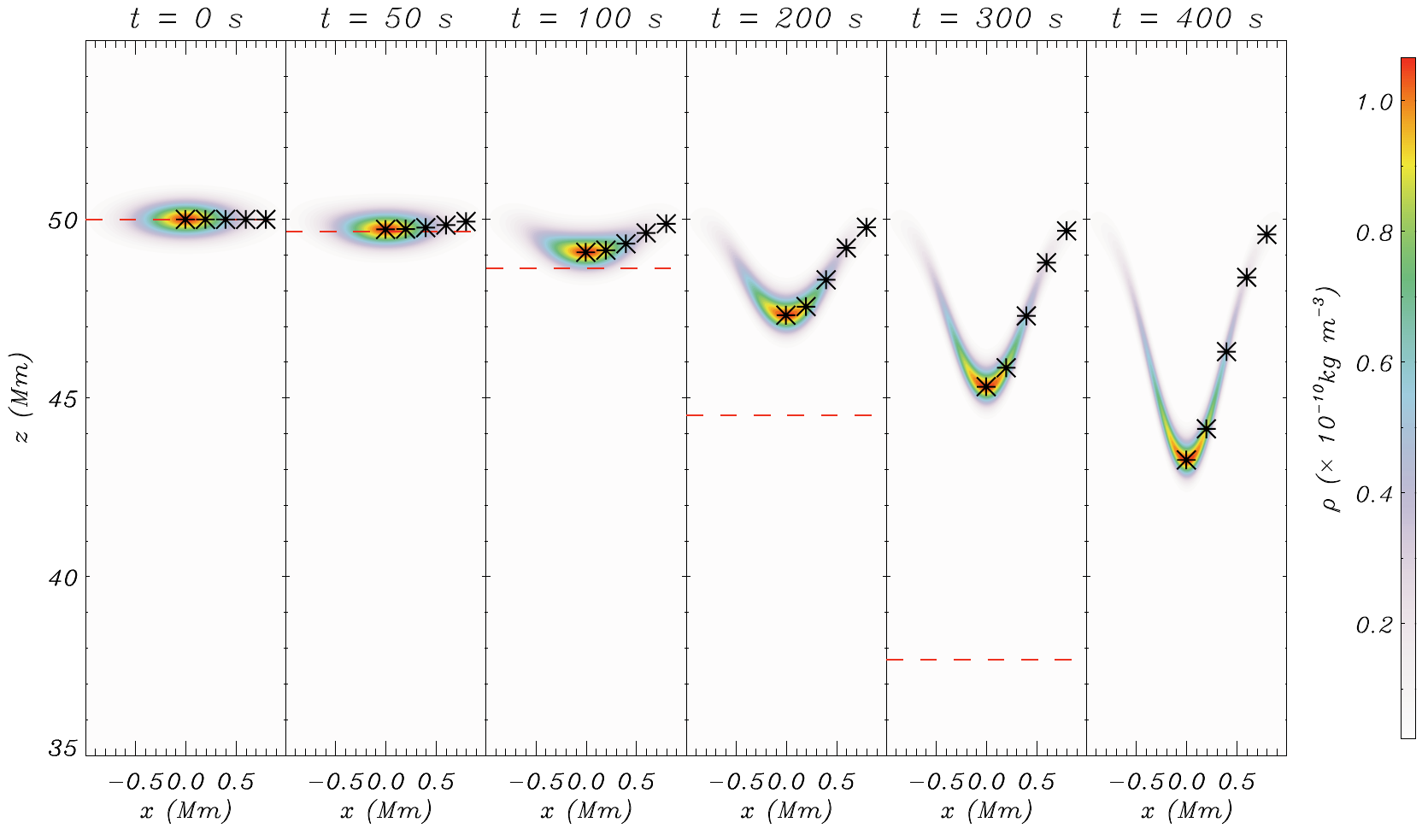}
    	\caption{Density at different times of the simulation I-5. The horizontal red dashed lines represent the position that would correspond to a free-falling object. The black asterisks show the position of the points whose kinematics is represented in Fig. \ref{fig:points}.
    	(An animation of this figure is available.)}
    	\label{fig:blob}
    \end{figure*}

	In the first place, we focus on the output of the simulation with the lowest values of the blob density and the magnetic field, i.e, that denoted as I-5. The height of the physical domain in this simulation is $L_{z} = 80 \ \rm{Mm}$, with the bottom boundary at $z_{\rm{bot}} = 0$.
	
	\begin{figure*}
		\centering
		\includegraphics[width=0.45\hsize]{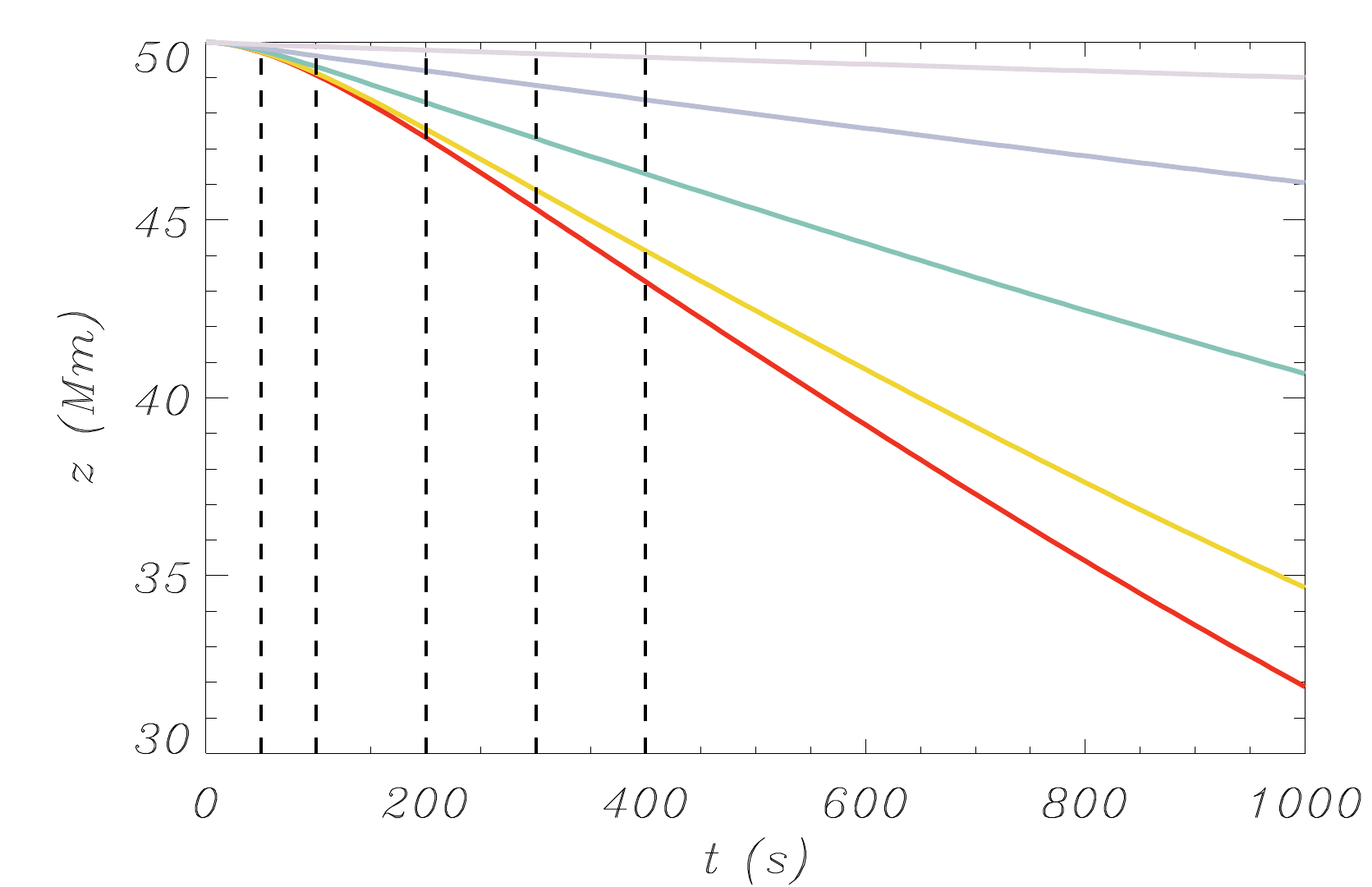} \includegraphics[width=0.45\hsize]{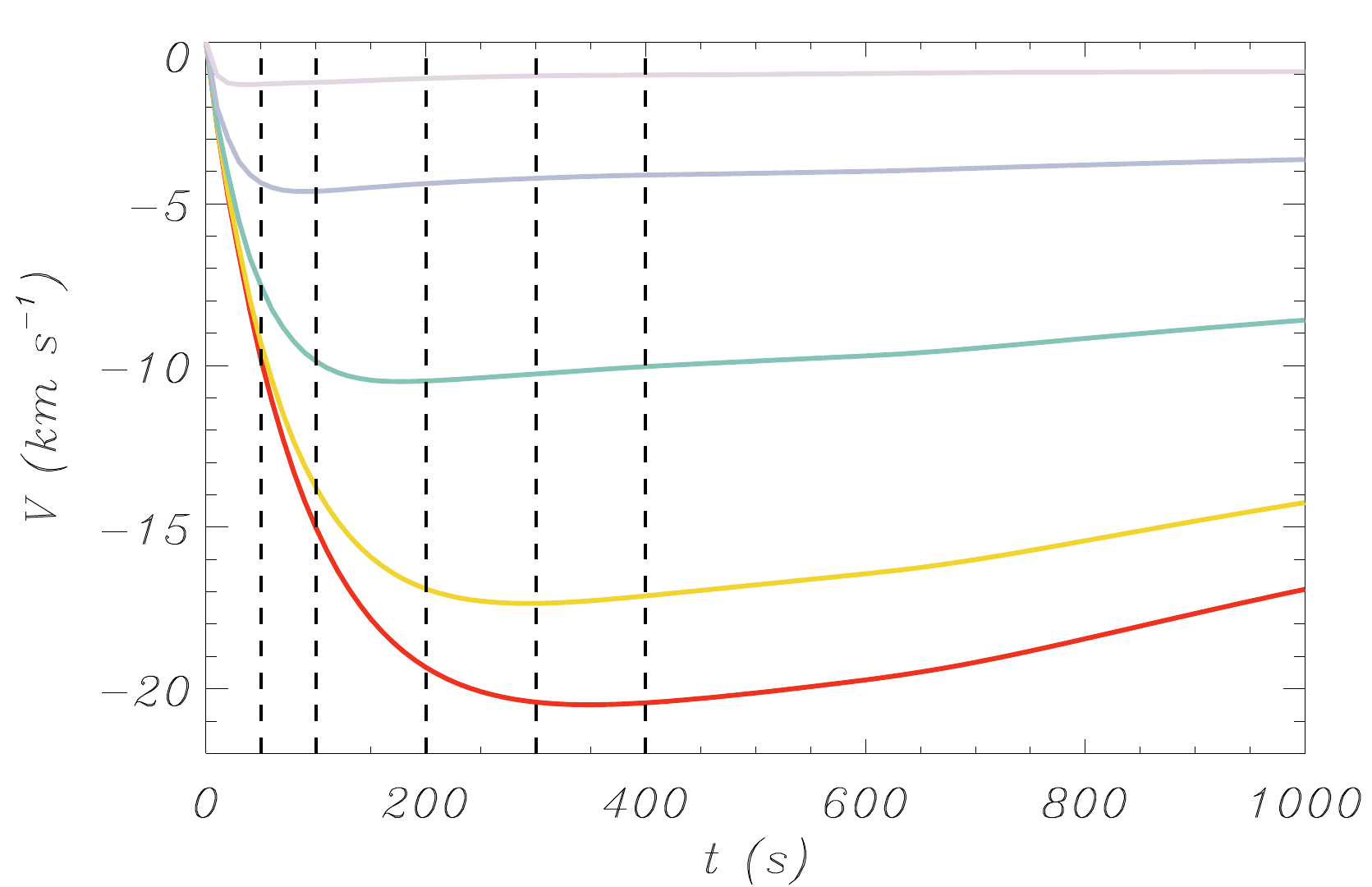}
		\caption{Left panel: position as a function of time of selected points from simulation I-5. Right panel: velocity as a function of time. Vertical dashed lines represent the time of the snapshots displayed in Fig. \ref{fig:blob}. The colours of the lines are related to the density scale shown in the previous figure.}
		\label{fig:points}
	\end{figure*}
	
	Figure \ref{fig:blob} displays several snapshots of the temporal evolution of the density. In addition, a horizontal red dashed line has been drawn to mark the position that the blob would have if it followed a free-fall motion. The blob is initially situated at the position $z_{0} = 50 \ \rm{Mm}$. At this height, the density ratio between the condensation and the surrounding hotter corona is $\sim 30$. Then, since the blob is not in mechanical equilibrium, it starts to fall under the action of gravity. However, it can be clearly seen that its descent is much slower than what is expected for a free-falling object: as shown by the rightmost panel, after a time of $400 \ \rm{s}$ the central part of the blob has only fallen a distance of $\sim 7 \ \rm{Mm}$, while a free-falling plasma would be out of the limits of the plot.
	
	Another noteworthy feature can be appreciated in Fig. \ref{fig:blob}. It can be seen even at the initial instants of the simulation that the blob is not able to keep its original shape as it falls. The central denser part moves faster than the lighter sides, which leads to the development of a characteristic V-shape. We explain later the reason why this shape appears.
	
	To better understand the behaviour described in the previous paragraphs, we have selected different points along the horizontal axis of the blob and investigated their kinematics during the whole simulation. These points are marked as black asterisks in Fig. \ref{fig:blob}. The results of this study are shown in Fig. \ref{fig:points}, which includes plots of the position and falling speed of the selected points as functions of time. All the points start at the same height, but as time advances, it can be seen that those associated with a smaller density start to lag behind the denser ones. 
	
	The different lines in the plots of position and falling speed show that, qualitatively, all the points have the same behaviour: at first, they are found in an acceleration phase, that is then followed by a stage in which the falling speeds become constant and even suffer a small decrease. The disparities appear in the duration of the acceleration phase, the values of the maximum falling speeds and the heights reached by each point. It can be checked that larger densities lead to longer acceleration phases and larger falling speeds. As a consequence, the points with a larger density reach lower heights faster. The two-stage behaviour represented by the plots of position as a function of time is in good agreement with the time-distance diagrams obtained from observations of coronal rain events \citep{2004A&A...415.1141D,2010ApJ...716..154A}.
	
\begin{figure}
		\centering
		\resizebox{\hsize}{!}{\includegraphics[]{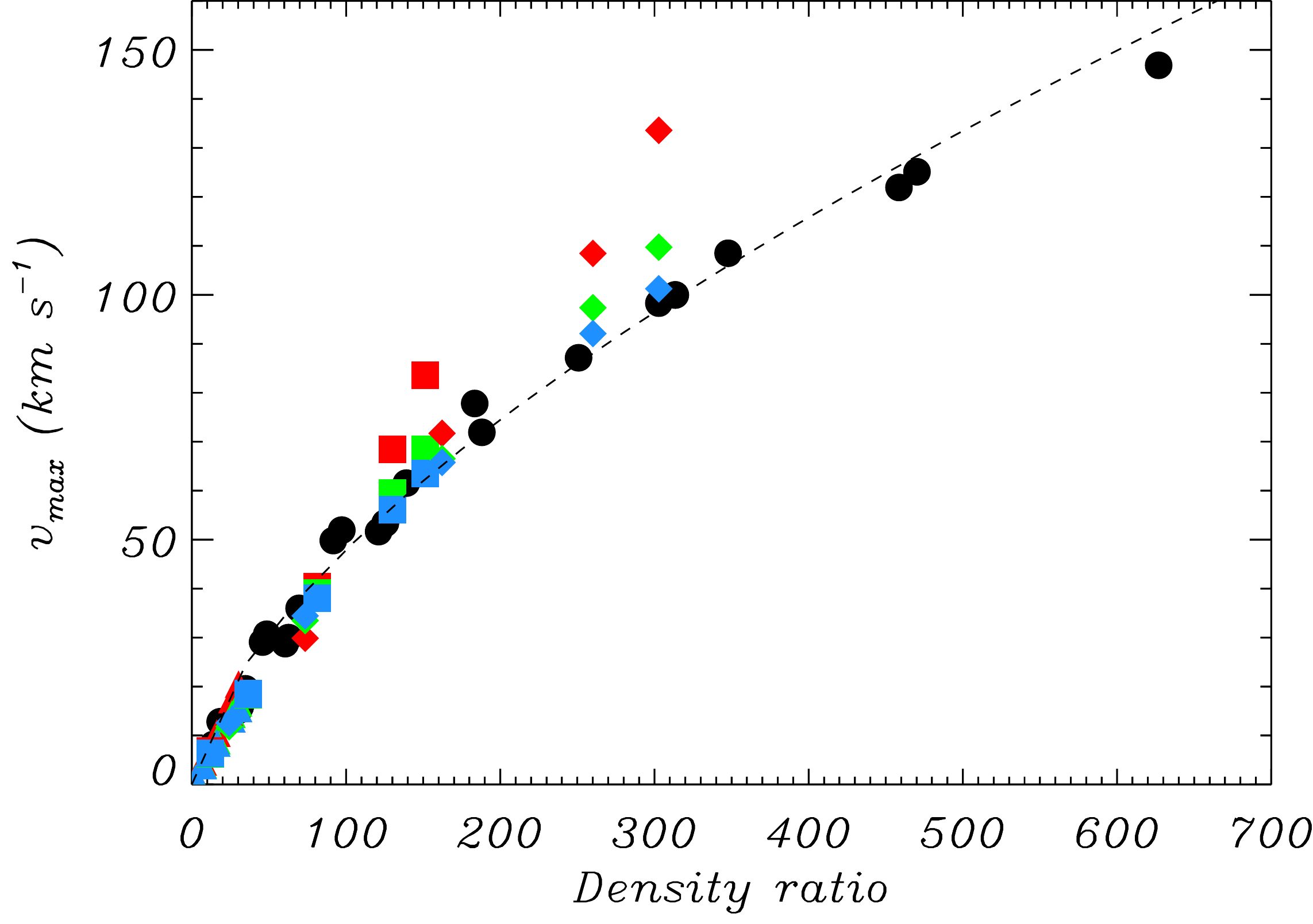}}
		\caption{Maximum falling speed as a function of density ratio. Black circles correspond to the results from 1D simulations. Triangles, squares and diamonds show the results of 2D simulations with $\rho_{\rm{b0}} = 1, \ 5 \ \rm{and} \ 10 \times 10^{-10} \ \rm{kg \ m^{-3}}$, respectively. Red, green and blue colour correspond to the cases with $B_{0} = 5, \ 10 \ \rm{and} \ 20 \ \rm{G}$, respectively.}
		\label{fig:ratio}
	\end{figure}

	To better study the relation between the density of the descending plasma and its maximum falling speed we have also performed simulations with larger values of the peak density: $\rho_{\rm{b0}} = 5 \times 10^{-10} \ \rm{kg \ m^{-3}}$ and $\rho_{\rm{b0}} = 10^{-9} \ \rm{kg \ m^{-3}}$, which correspond to density ratios at the blob peak of $\sim 150$ and $\sim 300$, respectively. For these simulations, the height of the domain is $L_{z} = 100 \ \rm{Mm}$, starting at $z_{\rm{bot}} = -40 \ \rm{Mm}$. The reason for using this different height is that the blob falls much faster in these simulations than in the previous one and arrives at positions below $z = 0$ (this means that, in a more realistic model, the blob would impact against the chromosphere). We have also repeated all these simulations but increasing the magnetic field strength to 10 and 20 G, and we have again selected different points of the condensation to investigate their kinematics. The results of this research are illustrated by Fig. \ref{fig:ratio}, where the maximum falling speed is represented as a function of the density ratio between the plasma of the blob and the surrounding corona at $t = 0$. The black circles show the results obtained from the 1D simulations analysed in \cite{2014ApJ...784...21O,2016ApJ...818..128O}. The remaining symbols correspond to the 2D simulations described in the present work: triangles, squares and diamonds represent the cases with $\rho_{\rm{b0}} = 1, \ 5, \ 10 \times 10^{-10} \ \rm{kg \ m^{-3}}$, respectively, and red, green and blue correspond to simulations with $B_{0} = 5, \ 10, \ 20 \ \rm{G}$, respectively (the specific colour and symbol for each simulation is shown in Table \ref{table:sims}).

	A clear trend is found in Fig. \ref{fig:ratio}: an increase in the density ratio leads to a larger falling speed. The black dashed line represents the fitting of the numerical data to the analytical expression
	\begin{equation}
		v_{\rm{max}} = a \Theta^{b},
	\end{equation}
	where $\Theta$ is the density ratio. We have found that the observed trend is well described by the fitting parameters $a = 2.56$ and $b = 0.64$. 
	
	It is remarkable the good agreement between the 1D and the 2D results displayed in Fig. \ref{fig:ratio}. But it is also unexpected. Here we have used a two-dimensional instead of a one-dimensional approach to investigate coronal rain and have included some effects, like the presence of a magnetic field, that were not taken into account in \citet{2014ApJ...784...21O,2016ApJ...818..128O}. So, why is there such a good agreement in the results in this large range of parameters when the models are, in principle, clearly different?
	
	Figure \ref{fig:ratio} also shows that there are some points corresponding to 2D simulations that do not conform to the overall trend. These ``misbehaving'' points are obtained from simulations with large densities and small values of magnetic field, while those with higher values of the initial magnetic field strength follow well the 1D general trend.
	
	The results presented in Figs. \ref{fig:blob} -- \ref{fig:ratio} allow us to describe the motion of the falling blobs and draw a relation between some parameters but they do not explain the reasons of the observed behaviours. For instance, up to this point of the research we can say that for small enough magnetic fields the 2D results depart from the 1D predictions. On the other hand, the agreement between the 1D and 2D results for larger values of $B_{0}$ indicates that the dynamics of each vertical slice $x$ = const. is independent of the neighbouring slices. To explain all this we need to look into the dynamics of the system. This is done in the following section.

\subsection{Blob dynamics} \label{sec:dyn}
    From the equation of momentum conservation we can determine the different accelerations the plasma is subject to during its descending motion. Equation (\ref{eq:momen}) can be recast as
	\begin{equation} \label{eq:accel}
		\frac{\partial \bm{V}}{\partial t}=-\bm{V} \cdot \nabla \bm{V}-\frac{\nabla P}{\rho}+\frac1{\rho}\bm{J} \times \bm{B}+\bm{g},
	\end{equation}
    where $\bm{J}=\frac{\nabla \times \bm{B}}{\mu_{0}}$ is the current density. In this form, we can identify the first term on the right-hand side as the inertial term and the second one as the acceleration due to the pressure gradient. The third one is the Lorentz term related to the presence of the magnetic field and the last one is the acceleration due to gravity.
    
\begin{figure}
    	\centering
    	\resizebox{\hsize}{!}{\includegraphics[]{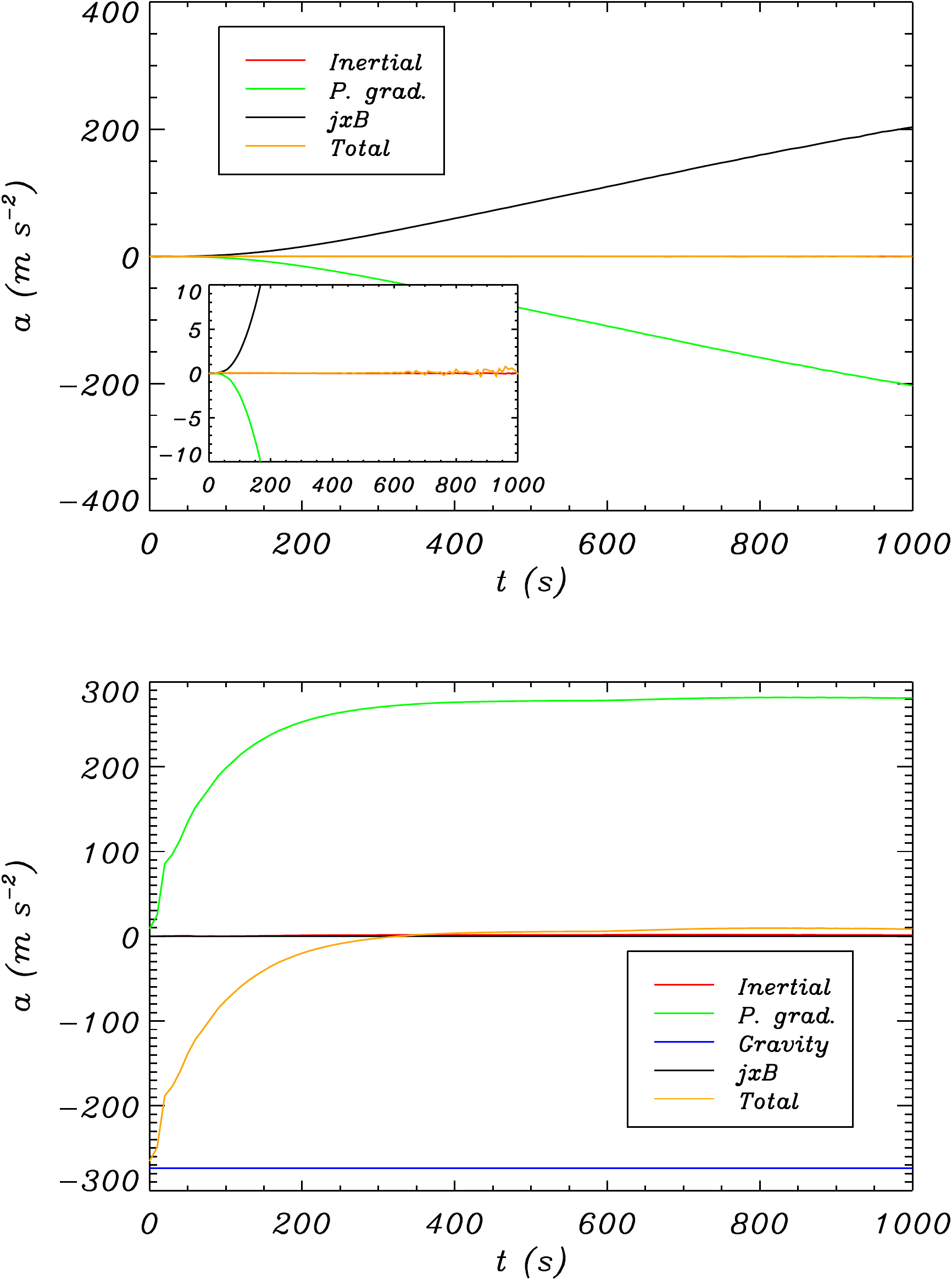}}
    	\caption{Dynamics of the density peak in the simulation I-5. Top: accelerations in the $x$-direction. Bottom: acceleration in the $z$-direction.}
    	\label{fig:accel}
    \end{figure}

    In Fig. \ref{fig:accel} we have represented the different components of the acceleration at the density peak of the condensation as functions of time, i.e., following the motion of the black cross at $x = 0$ in Fig. \ref{fig:blob}. The results are taken from simulation I-5. The top panel displays the accelerations along the horizontal direction, while the bottom panel shows the accelerations in the vertical direction. 

     First, we focus on the horizontal accelerations. The top panel of Fig. \ref{fig:accel} shows that the inertial component of the acceleration (represented by the solid red line) is negligible and we do not have an acceleration due to gravity because it is directed along the $z$-direction. Thus, the dynamics in the horizontal axis is mainly determined by the pressure gradient (green line) and the Lorentz term (black line). In absolute values, both terms reach very large magnitudes. Nevertheless, they are oriented in opposite directions and they balance each other almost perfectly, which leads to a total horizontal acceleration (orange line) that is almost zero: as shown in the inset, it varies in the range $[-1, 1] \ \rm{m \ s^{-2}}$. The same fact is observed at different points away from the $x = 0$ axis. Horizontal accelerations at these out-of-axis points are slightly larger than those for $x = 0$, with absolute values of up to $5 \ \rm{m \ s^{-2}}$, which are still negligible in comparison to the vertical accelerations. This means that the motion of the blob is mainly directed along the $z$-axis.
     
     On the other hand, the bottom panel of Fig. \ref{fig:accel} shows that the inertial term and the Lorentz term are both negligible in the vertical direction. Consequently, the descending motion depends on the relation between the pressure gradient and gravity. At $t = 0$, there is no pressure gradient and the only acting force is gravity. Thus, plasma starts falling with an acceleration of $\sim 274 \ \rm{m \ s^{-2}}$. But immediately a pressure gradient starts to build up, which opposes the action of gravity. This explains the acceleration phase seen in the observations and explains why the condensation does not follow a free-fall motion. The same conclusion was reached in \citet{2014ApJ...784...21O,2016ApJ...818..128O}. As time advances, the pressure gradient keeps increasing until it completely balances the action of gravity, which would correspond to the constant velocity phase. According to the plot, this occurs at a time around $t = 200 \ \rm{s}$. From this time on, it can be seen that the total vertical acceleration is not exactly zero but becomes positive and, in the last steps of the simulation, reaches values of up to $10 \ \rm{m \ s^{-2}}$. This means that the total acceleration points upwards and the falling speed is slightly reduced, in agreement with the velocity profiles shown in Fig. \ref{fig:points}. 
     
     The time it takes the pressure gradient to balance gravity, i.e., the duration of the acceleration phase, depends on the density, a dependence that is then reflected in the maximum falling speed reached by the plasma, as shown in Fig. \ref{fig:ratio}. According to Eq. (\ref{eq:accel}), the acceleration related to the pressure gradient is inversely proportional to the density. So, points of larger density require a larger pressure gradient to compensate the acceleration of gravity (see Appendix \ref{sec:app} for more details on the pressure gradient). The fact that horizontal forces cancel out implies that each vertical slice $x = const.$ does not interact with its neighbours, so that its dynamics is only governed by the density ratio at this value of $x$. This explains why the central part of the blob falls faster than its sides and the V-shape develops: it happens due to its horizontal variation of density.

	\begin{figure*}
		\centering
		\resizebox{\hsize}{!}{
		\includegraphics[]{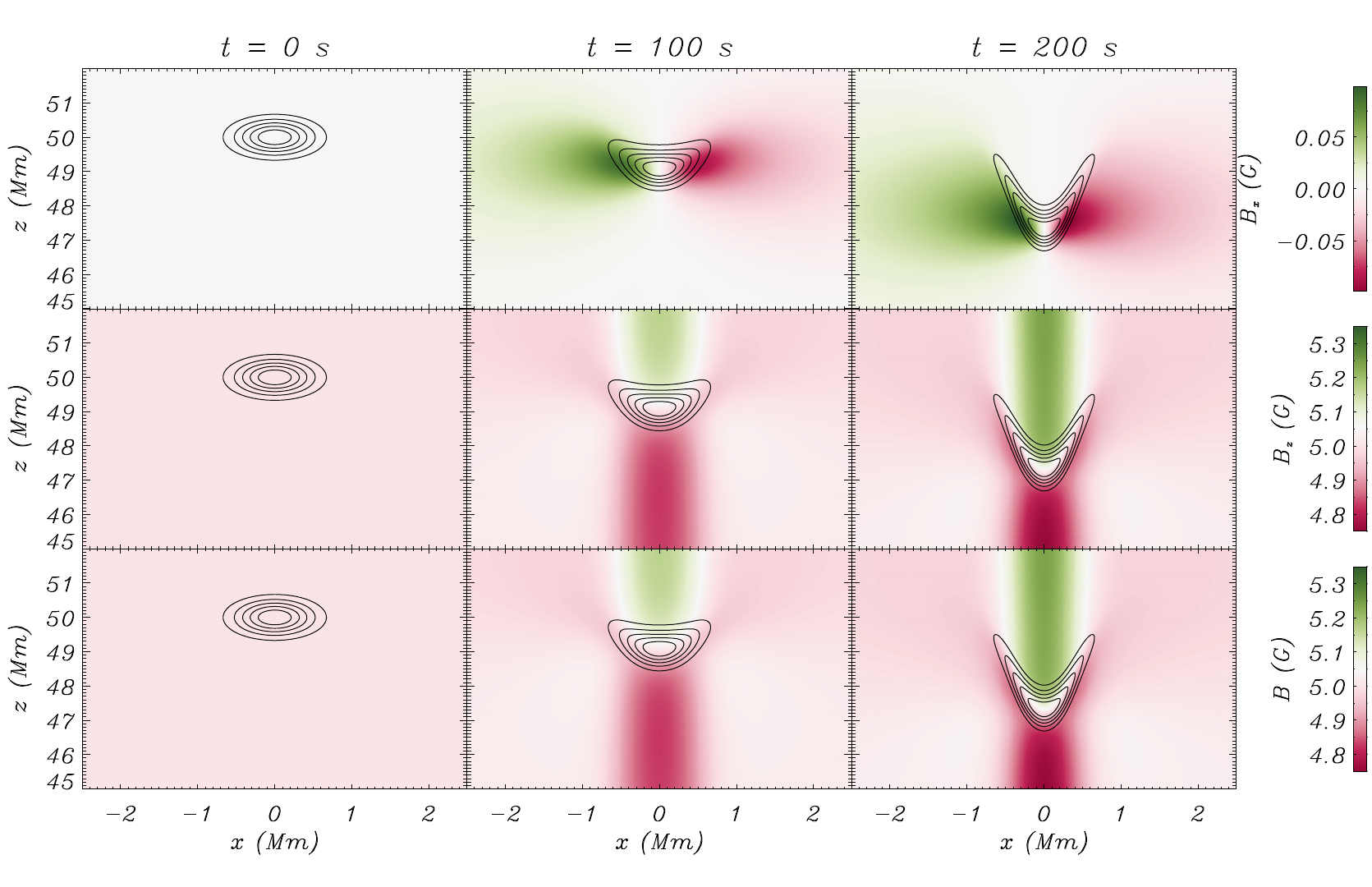}}
		
		\caption{Colour maps of $B_{x}$ (top), $B_{z}$ (middle) and $B$ (bottom) from simulation I-5. Black lines are density contours.}
		\label{fig:2D_bfield}
	\end{figure*}

	The analysis detailed in the paragraphs above focused on specific points in the numerical domain of the simulations. Now, we want to give a broader perspective of the dynamics. For this reason, in Fig. \ref{fig:2D_bfield} we display 2D maps of the $x$-component (top), $z$-component (middle) and absolute value (bottom) of the magnetic field at three different times of the same simulation discussed before. The purple and green colours represent the strength of the magnetic field and the black lines are contours of density that show the position and shape of the condensation at each time.
	
	The left column of Fig. \ref{fig:2D_bfield} shows that initially the magnetic field does not have a component in the $x$-direction, but it has only a $z$-component, with a uniform strength of $B_{0} = 5 \ \rm{G}$. As time advances and the plasma starts to fall, perturbations appear in both components. The amplitude of these perturbations is less than $5 \%$ of the initial magnetic field strength for the case of $B_{z}$, and even less for $B_{x}$. 	
    The middle and bottom panels show that the perturbations mainly propagate in the vertical direction and form a column in front of and behind the descending blob, while the rest of the domain remains almost unaffected. However, there is also a small spread along the horizontal direction, which is more evident in the top panels.
	
\begin{figure*}
    	\centering
    	\resizebox{\hsize}{!}{
    		\includegraphics[]{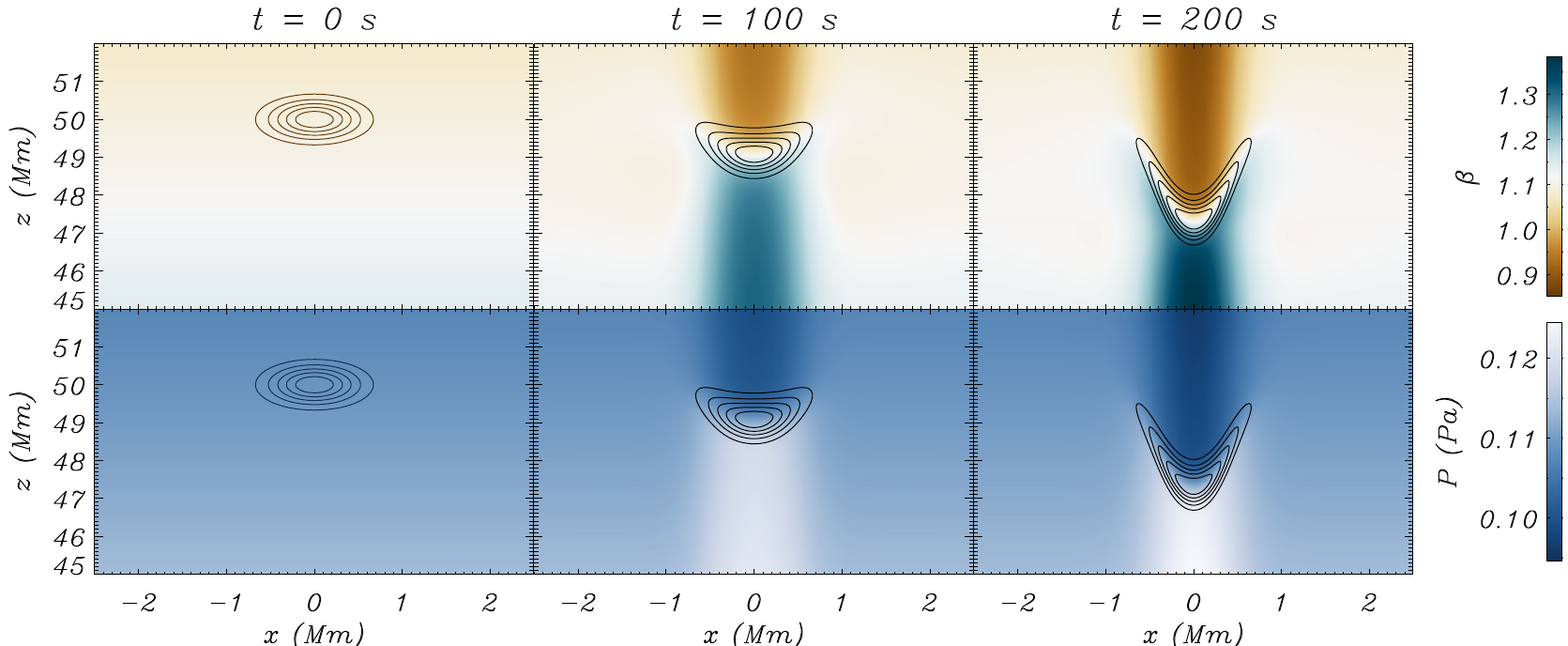}}
    	
    	\caption{Colour maps of plasma-$\beta$ (top) and pressure (bottom) from simulation I-5. Black lines are density contours.}
    	\label{fig:2D_pressure}
    \end{figure*}

	A similar behaviour is found in other variables, as shown by Fig. \ref{fig:2D_pressure}, where we have represented the plasma-$\beta$ (top) and the pressure (bottom). The black lines correspond to density contours. Again the perturbations are seen to propagate mainly along a vertical column whose horizontal extension is given by the size of the blob. Although the plasma-$\beta$ is larger at the front of the blob than at its back, its magnitude stays around unity, which means that neither the magnetic field nor the gas dynamics completely dominates the motion of the plasma.

	These results correspond to the case of small peak density and small magnetic field. But, as mentioned before, we have also performed simulations with larger values of these parameters. Since $\beta$ is proportional to the gas pressure (which is proportional to the density) but is inversely proportional to the square of the magnetic field strength, we expect very different values of $\beta$ in these simulations. Increasing the density yields an increase of $\beta$, meaning that the gas dynamics becomes more important than the magnetic field. However, an increase of the magnetic field produces a much smaller $\beta$ and, therefore, the motion of the plasma becomes more dominated by the magnetic field.
	
\begin{figure}
		\centering
		\resizebox{\hsize}{!}{\includegraphics[]{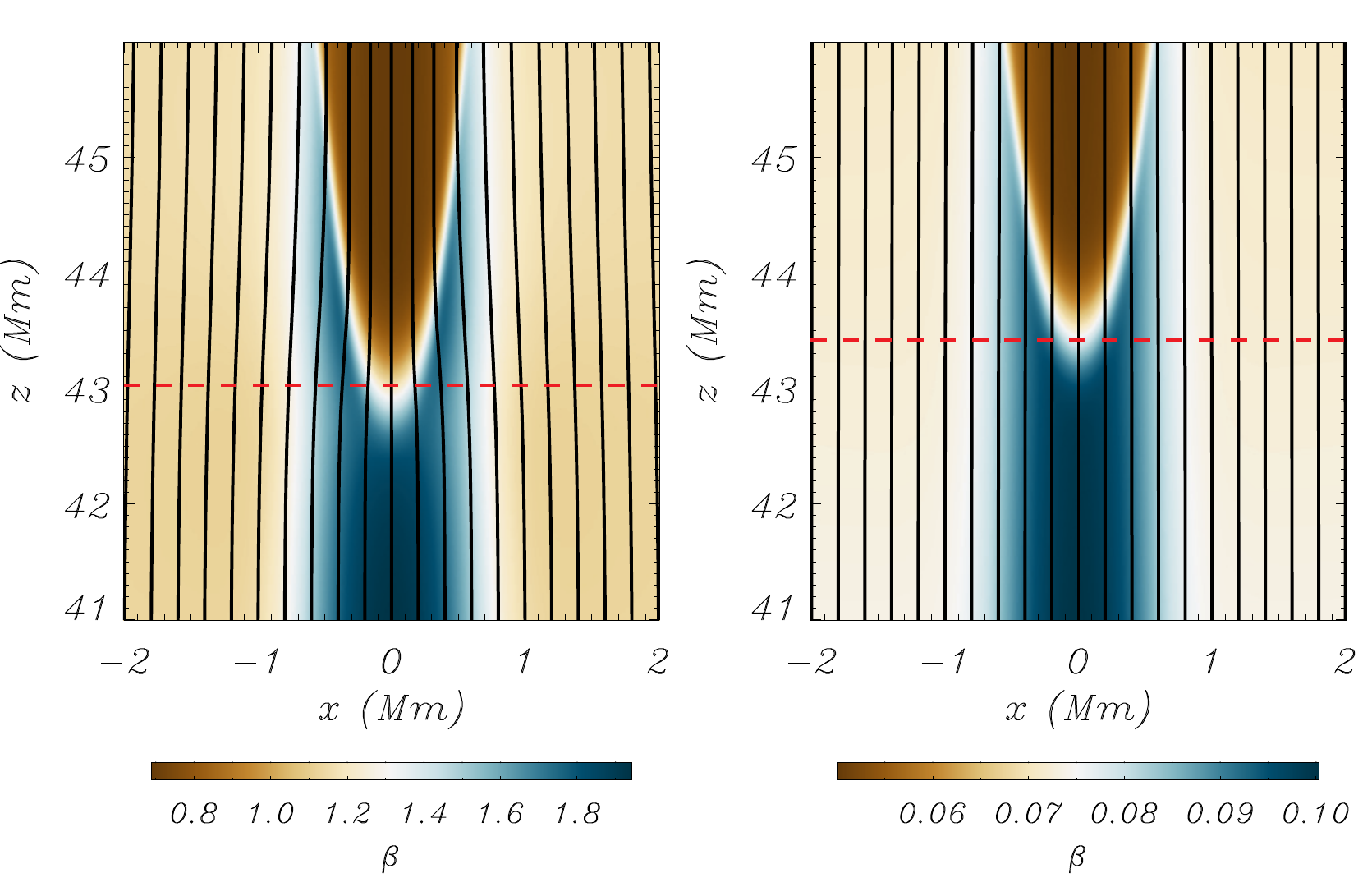}}
		\caption{Snapshots from simulations V-5 (left) and V-20 (right) at time $t = 250 \ \rm{s}$. The colour maps represent the plasma-$\beta$, magnetic field lines are displayed in black and the horizontal red dashed lines mark the height of the density peak of the blob.}
		\label{fig:lines}
	\end{figure}

	In Fig. \ref{fig:lines} we show snapshots from simulations with a larger density than before. The panel on the left corresponds to the simulation V-5 while the panel on the right corresponds to the simulation V-20. The colour map represents the plasma-$\beta$ and the black lines are the magnetic field lines. The position of the density peak is marked by the horizontal red dashed lines. The left panel, associated with $\beta > 1$, shows that the field lines are not completely vertical but they are distorted and have some curvature around the position of the blob. On the other hand, for the case of a stronger magnetic field, the lines are almost completely straight. They are hardly distorted by the presence of the plasma. Moreover, it can be seen that the blob is not at the same height in both simulations although the snapshots correspond to the same time. The plasma has fallen faster in the case with a smaller magnetic field.
	
	Now, we can fully understand the results summarised by Fig. \ref{fig:ratio}. The effect of the magnetic field is to try to confine the motion of the plasma into the vertical direction and reduce or prevent the motions along the horizontal direction. With a strong enough magnetic field, each plasma element falls following the path determined by a field line, with no influence from the adjacent elements. In addition, perturbations such as that of pressure or velocity can only propagate along the same line. Thus, the motion of each vertical cut of the descending blob can be described in terms of the 1D dynamics analysed in \citet{2014ApJ...784...21O,2016ApJ...818..128O}. On the other hand, with a weaker $B_{0}$, the dynamics is not completely dominated by the magnetic field and the perturbations can spread along the horizontal direction. This leads to a smaller pressure gradient opposing the action of gravity, which results in a longer acceleration phase and a larger maximum falling speed. Thus, we have found the explanation for the outlier points in Fig. \ref{fig:ratio}.
	
\subsection{Kelvin-Helmholtz instability} \label{sec:inst}
	All the simulations analysed in the previous section correspond to cases in which the magnetic field has a considerable contribution to the dynamics of the descending condensations. So, up to this point, we have left unexplored the regime in which the influence of the magnetic field is much smaller or can be completely neglected. Now, we want to investigate this regime. In this section we start by analysing the results of a simulation in which no magnetic field has been included in the equilibrium state. Therefore, this simulation is purely hydrodynamic. Then, we turn back to the magneothydrodynamic case but with a weak magnetic field.
	
	\begin{figure*}
		\centering
		\includegraphics[width=17cm]{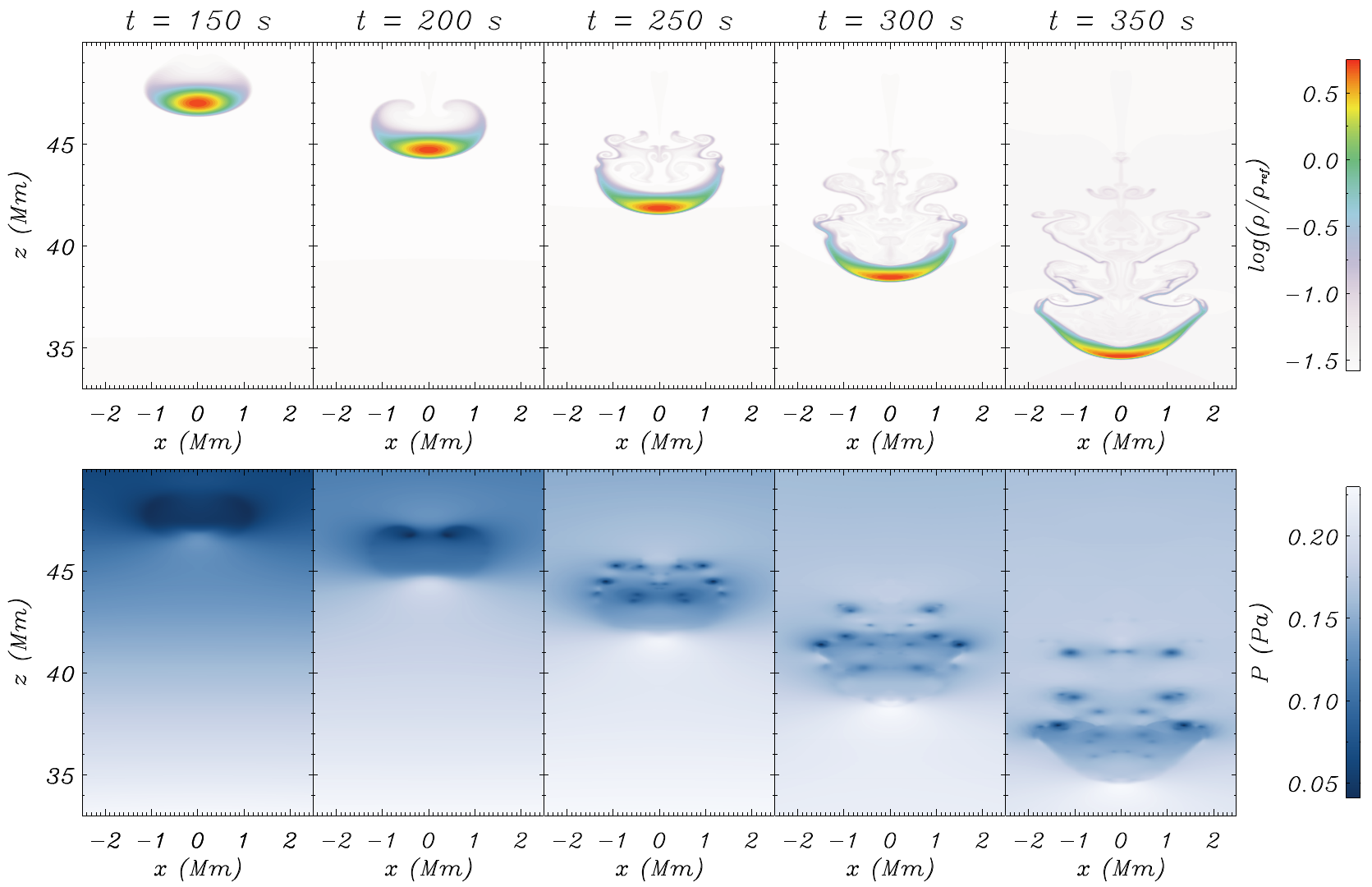}
		\caption{Colour maps of density (top) and pressure (bottom) from simulation V-0. (An animation of this figure is available.)}
		\label{fig:khi2D}
	\end{figure*}
	
	Figure \ref{fig:khi2D} displays several snapshots of the simulation V-0. Except for the magnetic field, it has the same parameters as those used to generate Fig. \ref{fig:lines}. The top and bottom panels show the temporal evolution of density and pressure, respectively. Looking at the top panels, it can be seen that some ripples appear at the sides of the blob as it falls. As time advances, these ripples turn into vortexes, features with smaller scales appear and the shape of the blob becomes much more distorted than in the cases that took into account the presence of the magnetic field. This behaviour reminds of the one expected when the conditions for the development of the Kelvin-Helmholtz instability (KHI) are fulfilled. Another remarkable fact can be seen in the bottom panels: in contrast with the magnetohydrodynamic simulations analysed before, here the perturbations of pressure clearly propagate in all directions and are not confined to a vertical column. 
	
	The KHI is a phenomenon that occurs when there is a shear flow at the interface between two different fluids. However, the existence of the shear flow does not always guarantee the emergence of this instability. There are some physical effects, such as gravity or the surface tension of the fluids, that prevent its development or greatly reduce its growth rate \citep{1961hhs..book.....C}.
	
	The most simple way to study the KHI is to consider two fluids of different but uniform densities, $\rho_{1}$ and $\rho_{2}$, and a shear flow represented by a discontinuous jump at the interface between the two fluids. In this configuration, according to \citet{1961hhs..book.....C}, the presence of a magnetic field oriented in the same direction as the shear flow has an stabilising effect on fully ionised plasmas. Then, the plasma only becomes unstable if the velocity of the flow is larger than a certain threshold. The value of this threshold is given by
    \begin{equation} \label{eq:threshold}
    	V_{\rm{KHI}} = \sqrt{\frac{B_{1}^{2}+B_{2}^{2}}{\mu_{0}}\frac{\rho_{1}+\rho_{2}}{\rho_{1}\rho_{2}}},
    \end{equation}
    which assumes that the magnetic field felt by each fluid may be different.
    
    In the simulations analysed in Sections \ref{sec:kin} and \ref{sec:dyn}, we have found no signs of the development of the KHI. The smallest value of the magnetic field strength used in that series of simulations was $B_{0} = 5 \ \rm{G}$, which may produce a large enough velocity threshold.
    
    To compute an approximate expression for the velocity threshold associated with that $B_{0}$, we set $B_{1}=B_{2}=B_{0}$, because we have a uniform magnetic field over all the domain, and $\rho_{1}=\rho_{\rm{eq}}$ and $\rho_{2} = \rho_{\rm{b0}}$. We also take into account that the equilibrium density varies with height but we assume that each height is an independent layer not affected by the layers above or below, i.e., we neglect the effects related to the vertical gradient of density. Thus, the velocity threshold can be expressed as
    \begin{equation}
    	V_{\rm{KHI}}(z) = \sqrt{\frac{2B_{0}^{2}}{\mu_{0}}\frac{\rho_{\rm{eq}}(z)+\rho_{\rm{b0}}}{\rho_{\rm{eq}}(z)\rho_{\rm{b0}}}}.
    \end{equation}
    Moreover, due to the density of the blob being much larger than the density of the surrounding corona, we can obtain the following approximate expression:
    \begin{equation} \label{eq:vkhi}
    	V_{\rm{KHI}}(z) \approx \sqrt{2}c_{\rm{Ac}}(z),
    \end{equation}
    where $c_{\rm{Ac}}(z)=B_{0}/\sqrt{\mu_{0}\rho_{\rm{eq}}(z)}$ is the Alfvén speed of the corona. This expression tells us that super-Alfvénic speeds are needed for the development of the KHI. Note that in this approximation the velocity threshold does not depend on the density of the blob.
    
    The work of \citet{1961hhs..book.....C} also provides the growth rate of the instability, which in our case can be written as
    \begin{equation} \label{eq:growth}
    	\omega_{\rm{KHI}}(z)=k_{z}\left[\frac{\rho_{\rm{eq}}(z)}{\rho_{\rm{b0}}}\left(V_{\rm{KHI}}(z)^{2}-\Delta U^{2}\right)\right]^{1/2},
    \end{equation}
    where $k_{z}$ is the wavenumber associated with the perturbation that may become unstable and $\Delta U$ is the shear flow velocity.
        
    Now, to check if the KHI may develop during the fall of the plasma, we can compare the falling speeds of the blob with the velocity thresholds provided by the previous equations. For this comparison we define $V_{fall}$ as the $z$-component of the velocity at the density maximum of the blob, whose time dependence is shown for a particular simulation at the right panel of Fig. \ref{fig:points}. Since the plasma surrounding the blob remains approximately static, the shear flow velocity is equal to $|V_{fall}|$ and the KHI only emerges if $|V_{fall}| > V_{\rm{KHI}}$. 
    
\begin{figure}
    	\centering
    	\resizebox{\hsize}{!}{\includegraphics[]{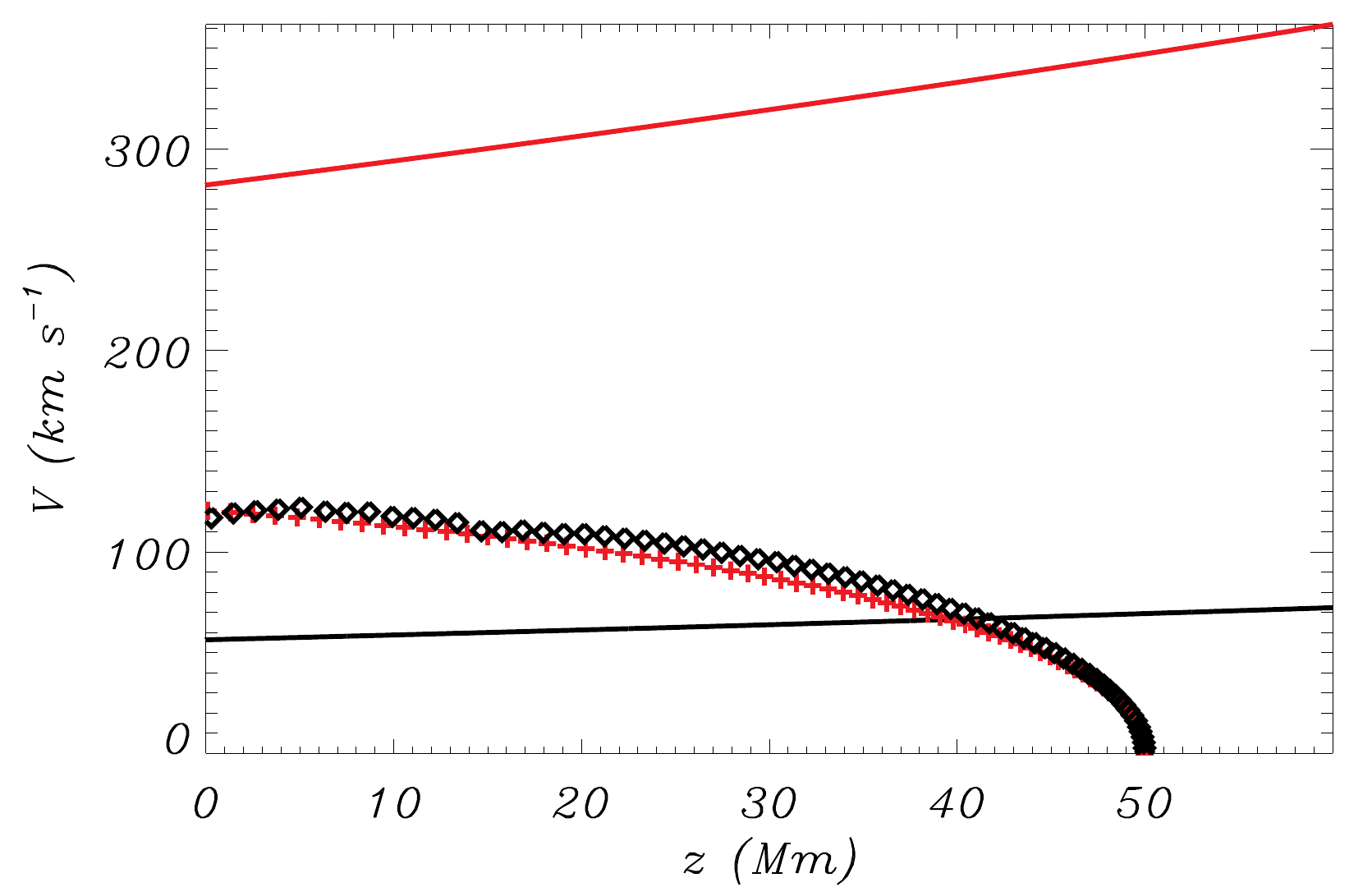}}
    	\caption{$|V_{fall}|$ and $V_{\rm{KHI}}$ as functions of height. Lines represent $V_{\rm{KHI}}$ and symbols represent the falling speed of blobs in simulations with $\rho_{b0} = 10^{-9} \ \rm{kg \ m^{-3}}$. Red colour corresponds to the case with $B_{0} = 5 \ \rm{G}$ while black corresponds to $B_{0} = 1 \ \rm{G}$.}
    	\label{fig:khi_height}
    \end{figure}

    In Fig. \ref{fig:khi_height}, we represent $V_{\rm{KHI}}$ as a function of height for two different magnetic field strengths: the red line corresponds to $B_{0} = 5 \ \rm{G}$ and the black line corresponds to $B_{0} = 1 \ \rm{G}$. We also display the values of $|V_{fall}|$ obtained from two different simulations with a large density blob: red crosses and black diamonds are the results from simulations X-5 and X-1, respectively.
    
    In the first place, we pay attention to the case with $B_{0} = 5 \ \rm{G}$. We can see that, although the blob reaches falling speeds of $\sim 120 \ \rm{km \ s^{-1}}$, $|V_{fall}|$ always stays well below $V_{\rm{KHI}}$, which has a minimum value of $\sim 280 \ \rm{km \ s^{-1}}$. This means that the KHI does not appear under these conditions. On the contrary, for $B_{0} = 1 \ \rm{G}$, the velocity threshold is small enough so that at a certain height $|V_{fall}|$ becomes larger than it. In this simulation, the condition for the development of the KHI is fulfilled once the blob falls to a height of $\sim 42 \ \rm{Mm}$. Then, the falling speeds keep increasing while the threshold reduces, which leads to larger growth rates of the instability.

    \begin{figure*}
    	\centering
    	\includegraphics[width=17cm]{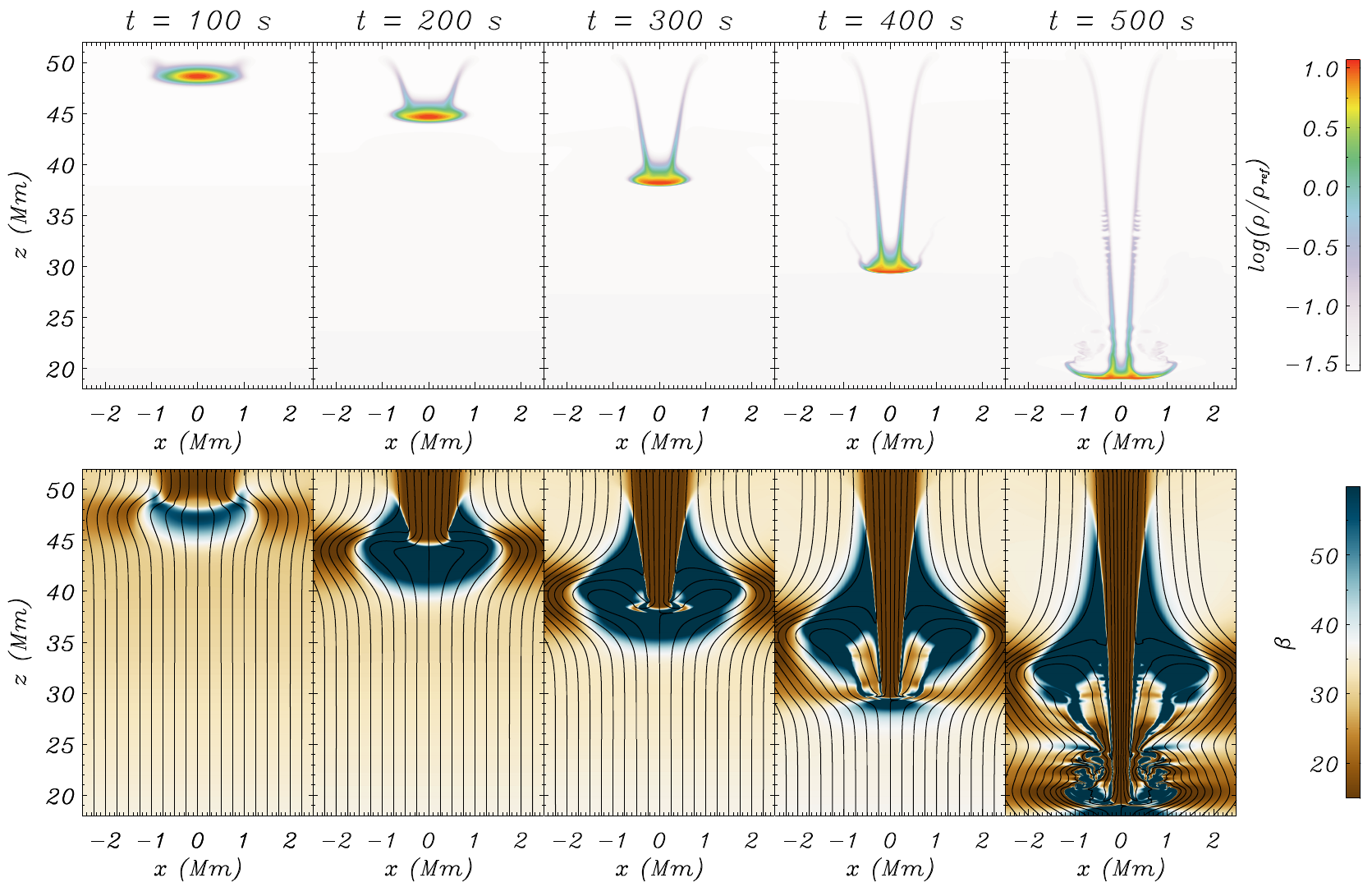}
    	\caption{Colour maps of density (top) and plasma-$\beta$ (bottom) from simulation X-1. Magnetic field lines are plotted in black colour in the bottom panel. (An animation of this figure is available.)}
    	\label{fig:khi2D_B01}
    \end{figure*}
	
	The direct comparison of the results from simulations X-5 and X-1 without taking into account their relation with the KHI thresholds may give the impression that both cases have a similar behaviour and that the KHI does not actually develop in the latter. The dependence of each respective $|V_{fall}|$ on height follows an almost identical trend. However, examining them in more detail, it can be checked that the condensation in X-1 falls slightly faster than in X-5. Moreover, we remind that $|V_{fall}|$ only provides information about the motion along the $x = 0$ axis. Therefore, we cannot infer the evolution of the whole condensation by only looking into this parameter.
	
	To check if the prediction from Fig. \ref{fig:khi_height} that the KHI should appear in simulation X-1 is correct, we need to examine its 2D behaviour. In Fig. \ref{fig:khi2D_B01} we display colour maps of density (top panels) and plasma-$\beta$ (bottom panels) at different snapshots of this simulation. The bottom panels also include a representation of the magnetic field lines. The panels for times $t = 100 \ \rm{s}$ and $t = 200 \ \rm{s}$ show the formation of the V-wings behind the blob. However, they do not emerge from the sides, as in Fig. \ref{fig:blob}, but from a more centered position. This is a consequence of the magnetic field being very weak, so it cannot prevent the horizontal motions as in the cases discussed in Sect. \ref{sec:kin}. The bottom panels show how the gas dynamics dominates and the magnetic field lines become very distorted. 
		
	Regarding the KHI, no sign of the ripples typically associated with this instability is seen in the leftmost panels of Fig. \ref{fig:khi2D_B01}. Nevertheless, the panels in the center, in which the condensation has already fallen below the height $z = 40 \ \rm{Mm}$, display a larger deformation of the blob. And at later times, the ripples and vortexes of the instability can be clearly seen. According to Fig. \ref{fig:khi_height}, the KHI may start to develop at heights below $z \sim 42 \ \rm{Mm}$, but since the growth rate of the instability depends on the difference between the falling speed and the velocity threshold, as shown by Eq. (\ref{eq:growth}), it would take some time until the perturbations on the plasma are large enough that they greatly alter the shape of the blob. Therefore, the results presented in Fig. \ref{fig:khi2D_B01} seem to agree with the prediction from Fig. \ref{fig:khi_height}.

	It must be noted that Fig. \ref{fig:khi_height} displays the results of two of the most unstable cases we have studied with our simulations. They have the largest blob densities of all the series, which lead to the largest falling speeds (see Fig. \ref{fig:ratio}), and the smallest magnetic field strengths, that correspond to the smallest velocity thresholds as shown by Eq. (\ref{eq:threshold}).
	
	In addition, this analysis has assumed a simplified model of the KHI, which considers that there is a discontinuous change of density and velocity between the two media and corresponds to the most unstable configuration. However, in our simulations we do not have that discontinuity but a continuous variation of density, which increases the stability of the system.
	
	Therefore, from our analysis we conclude that, due to the stabilising effect of the magnetic field, the coronal rain blobs are not expected to develop KH instabilities during their falling motion.

\section{Conclusions} \label{sec:concl}
    In the present work we have used a 2D numerical model to investigate the dynamics of descending condensations in the solar corona. We have assumed that the plasma is fully ionised and have not taken into account non-ideal effects (e.g., the ambipolar diffusion due to the presence of neutral particles or thermal conduction). We have represented the solar corona as an isothermal vertically stratified atmosphere with no boundaries, which means that we have neglected the effects related to the presence of the chromosphere and photosphere at the bottom boundary. Since we wanted to focus only on the descending phase of the coronal rain cycle, we did not include the formation process of the blobs in our study. Therefore, we have assumed that the condensations are already formed at the initial time of our simulations. We have represented them by a density enhancement described by a two-dimensional Gaussian profile. In addition, we have taken into account the presence of a uniform vertical magnetic field. In such configuration, the condensations are not in equilibrium and start to fall down under the action of gravity. In the present work we have analysed the features of this falling motion. 

	First, we have described the general evolution of the descending condensation, which does not behave as a free-falling object. The denser central parts fall faster than the lighter sides, which leads to the deformation of the initial Gaussian profile and the appearance of the V-shape displayed in Fig. \ref{fig:blob}. Then, we have studied the kinematics of several selected points along the horizontal axis of the condensation. Our numerical results show a good agreement with the time-distance diagrams elaborated from different observations of coronal rain events \citep{2001SoPh..198..325S,2004A&A...415.1141D,2010ApJ...716..154A}. Those diagrams reveal that the motion of the descending plasma can be separated in two phases: a first phase of acceleration and a second phase of almost constant falling speed.
	
	To understand the reasons behind the two phases of the descending motion and the development of the V-shape, we have analyzed the different forces that determine the evolution of the condensation. We have confirmed the findings of \citet{2014ApJ...784...21O} that the reason for the slower than free-fall descent is the appearance of a pressure gradient that opposes the action of gravity. As time advances, the pressure gradient builds up and, in some cases, is able to fully balance the acceleration due to gravity, which explains the existence of the constant speed phase. The magnitude of the pressure gradient required to balance gravity depends on the density of the plasma. Larger densities need larger pressure gradients, which leads to longer accelerations phases and larger maximum falling speeds. Therefore, the shape of a condensation with a horizontal variation of density will become distorted as the denser parts fall faster than the lighter ones. This explains the V-shape found in our simulations.
	
	The maximum falling speeds found in our study are below $150 \ \rm{km \ s^{-1}}$, which is in good agreement with the observations \citep{,1996SoPh..166...89W,2012ApJ...745..152A}. Moreover, our 2D results show a remarkable agreement with the 1D results from \citet{2014ApJ...784...21O,2016ApJ...818..128O}, except for the cases with a weak magnetic field. For weak fields, the maximum falling speeds are larger than in the 1D model. The analysis of the accelerations due to the presence of the magnetic field has shown that the effect of the Lorentz term in the vertical direction is negligible in comparison with those of the pressure gradient or gravity. On the other hand, it becomes fundamental in the motions along the horizontal direction. It almost perfectly balances the effect of the pressure gradient, leading to a negligible acceleration along the $x$-direction. As a consequence, the evolution of each vertical slice of the condensation becomes almost independent of its neighbours and its evolution can be described by the 1D model. Another effect of the presence of a strong enough magnetic field is that the propagation of the perturbations becomes confined to the vertical direction. For instance, the perturbation of pressure cannot spread along the $x$-direction, which results in larger vertical pressure gradient and, consequently, a smaller falling speed.
	
	We have also simulated cases in which the magnetic field strength is very small or there is no magnetic field at all. In the purely hydrodynamic configuration, i.e., when the effect of the magnetic field is neglected, the evolution of the descending blob is very different from that described in the previous paragraphs. As the plasma falls, ripples and vortexes related to the KHI appear and the shape of the blob is greatly distorted. In these circumstances, the 2D evolution of the condensation cannot longer be treated as a combination of independent vertical slices that follow the 1D dynamics.
	
	It is known that the presence of a magnetic field oriented along the direction of the shear flow has a stabilising effect on the KHI \citep{1961hhs..book.....C}. Thus, we have performed a study to determine under which conditions the KHI may develop during coronal rain events. Using a simple analytical model, we have compared the theoretical velocity thresholds for the emergence of the instability with the falling speeds computed from several simulations. Such comparison has shown that, for the majority of cases, the maximum falling speeds are well below the super-Alfvénic velocity thresholds. Only for very small values of the magnetic field strength the falling speeds are comparable to or larger than the theoretical thresholds. This means that under the typical conditions in which coronal rain occurs the KHI is not expected to develop.
	
	However, some circumstances that might modify the previous conclusion should be mentioned here. We have assumed a vertical uniform magnetic field. In a more realistic model, $B_{0}$ should have a decreasing dependence on height, which may lead to smaller values in the upper layers of the solar corona than the ones used in our simulations, thus reducing the instability velocity thresholds. So, it may be possible for the KHI to appear in blobs falling from higher layers of the corona.  Furthermore, it is known that the presence of neutrals in a plasma can have a dramatic effect on the development of the instability \citep{2004ApJ...608..274W}. The neutral component of the plasma is always unstable, even for sub-Alfvénic speeds, although the collisional coupling with ions reduces the growth rate of the instability \citep{2012ApJ...749..163S}. The influence of the dependence of the magnetic field on height and the presence of a neutral component in the condensation should be studied in the future.
	
	We should also remark that the V-shape of the blob found in our simulations does not agree with what has been actually detected. The picture we get from observations is that of an elongated condensation with no wings at its sides \citep{1996SoPh..166...89W,2012ApJ...745..152A}. This discrepancy may be explained by assuming that the catastrophic cooling leading to the formation of the blob is a continuous process that is still active while the plasma starts to fall. In this way, the coronal plasma can continue to condense and fall along the same path. On the contrary, our simulations start when the blob has already been completely formed and the cooling has already stopped. It is also possible that the wings are actually present in the coronal rain events but the instruments cannot detect them because their emission is much smaller than that of the central part of the blob, due to the great difference in density. Another alternative explanation comes from extrapolating our 2D results to a 3D scenario: if we assume that the blobs have axial symmetry, the 2D V-shape transforms into a funnel whose density decreases with height. This funnel-shape would have a greater resemblance to what the observations show. In any case, more complex simulations including the formation process of the condensations and forward modeling tools should be used to solve the discrepancy.
	
	Finally, as mentioned before, this is the first paper of a series. Hence, we have left out from the present study some effects that we would like to include in forthcoming works. For instance, we plan to perform simulations that take into account the influence of the chromosphere. The presence of this much denser layer can contribute to slowing down the blob descent and even produce a rebounding of the plasma, as shown by \citet{2001SoPh..198..289M}. In addition, here we have only considered a completely vertical magnetic field but observations show that the plasma in coronal rain moves along curved paths, following the magnetic structure of the coronal loops \citep{2006A&A...449..369T,2007A&A...472..633T,2012ApJ...745..152A,2012SoPh..280..457A}. We want to study the evolution of the condensation when the magnetic field lines are inclined or follow a curved path. In such configurations, we expect to find smaller falling speeds due to the effect of the projection of gravity along the path. The influence of the inclination or the curvature of the field lines may be enhanced when the condensation is assumed to be partially ionised. The neutral component of the plasma does not feel directly the guiding effect of the magnetic field and can drag the ionised component through the magnetic field lines. However, it is also possible that the collisional coupling between ions and neutrals is strong enough to prevent the latter from falling vertically. It is interesting to investigate how the dynamics of partially ionised blobs depends on the ionisation degree in addition to the inclination or curvature of the field lines.

\begin{acknowledgements}
	D.M., E.K., and M.C. acknowledge support from the European Research Council through the Consolidator Grant ERC-2017-CoG-771310-PI2FA. R.O. acknowledges support from the Spanish Ministry of Economy, Industry and Competitiveness through the grant AYA2017-85465-P. R.O. also acknowledges the travel support received from the International Space Science Institute (Bern, Switzerland) as well as discussions with members of the ISSI team on ``Observed multi-scale variability of coronal loops as a probe of coronal heating", led by C. Froment and P. Antolin. The authors thankfully acknowledge the technical expertise and assistance provided by the Spanish Supercomputing Network (Red Espanola de Supercomputacion), as well as the computer resources used: the LaPalma Supercomputer, located at the Instituto de Astrofisica de Canarias.
\end{acknowledgements}

\appendix 
\section{Profile of the pressure gradient} \label{sec:app}
	In Sect. \ref{sec:dyn}, we have shown that the inertial and the Lorentz terms of the momentum equation do not have an important effect on the vertical motion of the descending blob. Therefore, neglecting those terms, Eq. (\ref{eq:accel}) can be rewritten as
	\begin{equation}
		\frac{\partial V_{z}}{\partial t} = -\frac1{\rho}\frac{\partial P}{\partial z}-g.
	\end{equation}
	
	After the initial acceleration phase, the plasma keeps descending with a constant velocity, so $\partial V_{z} / \partial t = 0$ and $\nabla P = -\rho g$.
	
	The initial two-dimensional profile of the density enhancement is given by Eq. (\ref{eq:blob}), but if we only take now into account its dependence with height, we have that
	\begin{equation} \label{eq:blob_z}
		\rho_{\rm{b}}(z)=\rho_{\rm{b0}} \exp \left[-\left(\frac{z-z_{0}}{\Delta}\right)^{2}\right].
	\end{equation}
	
	Thus, the profile of the perturbation of pressure that balances the action of gravity is given by
	\begin{equation}
		P(z)=\int_{0}^{z}-g \rho \exp \left[-\left(\frac{z-z_{peak}}{\Delta}\right)^{2}\right] dz,
	\end{equation}
	which results in
	\begin{equation} \label{eq:p_z}
		P(z)= -\frac{g \rho_{\rm{b0}}}{2}\sqrt{\pi} \Delta erf \left(\frac{z-z_{peak}}{\Delta}\right),
	\end{equation}
	where $erf(z)$ is the Gauss error function. Note that here we have used $z_{peak}$ instead $z_{0}$ to take into account the actual position of the blob during its fall.
	
	Then, to obtain the total pressure required for a constant-speed phase, we have to combine the expression from Eq. (\ref{eq:p_z}) with the one corresponding to the pressure in the equilibrium state, given by Eq. (\ref{eq:equil}). Hence,
	\begin{equation} \label{eq:pc}
		P_{c}(z) = P_{0} e^{-z/H}-\frac{g \rho_{\rm{b0}}}{2}\sqrt{\pi} \Delta erf \left(\frac{z-z_{peak}}{\Delta}\right).
	\end{equation}
	
\begin{figure}
		\resizebox{\hsize}{!}{\includegraphics{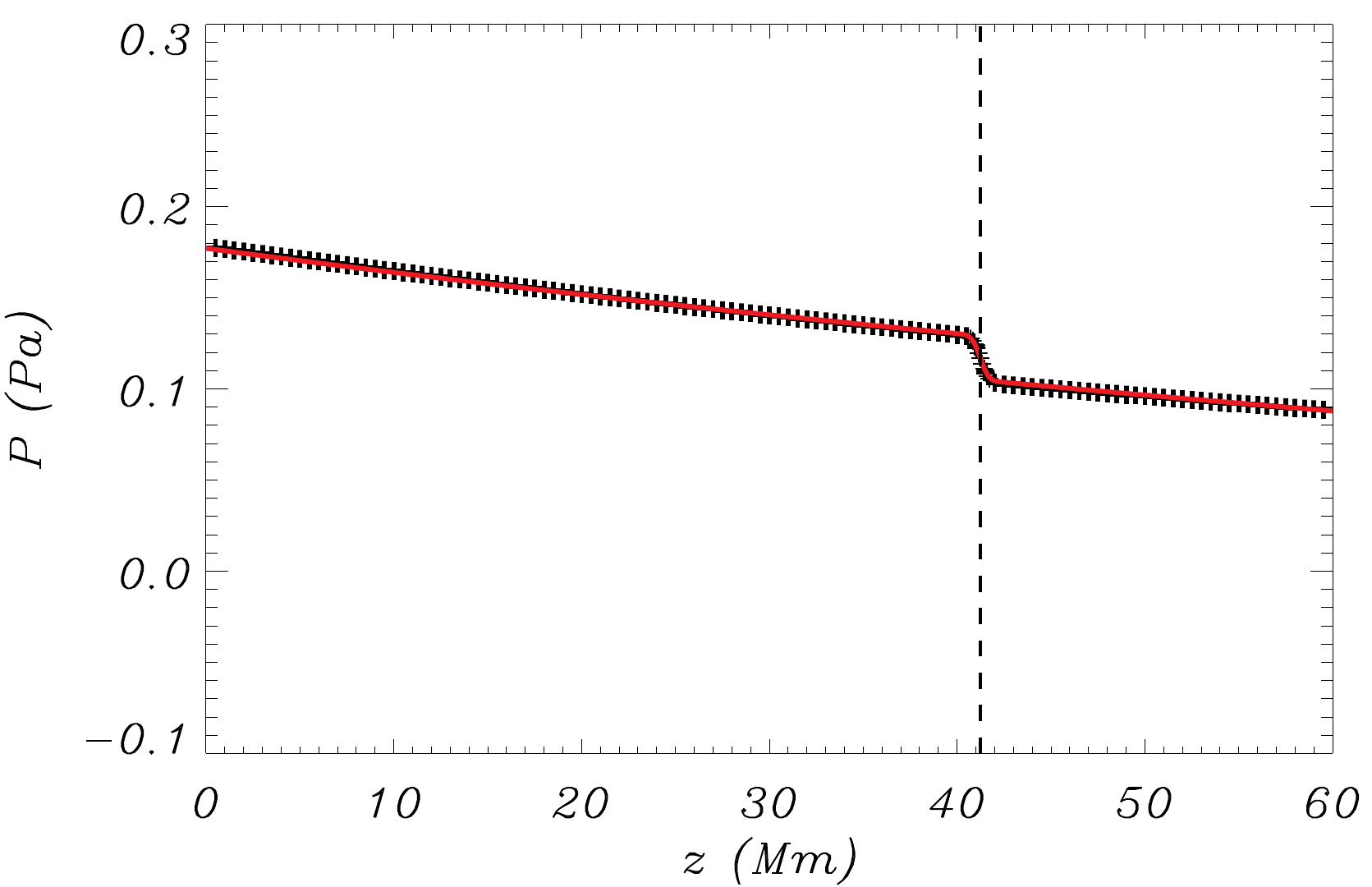}} \\
		\resizebox{\hsize}{!}{\includegraphics{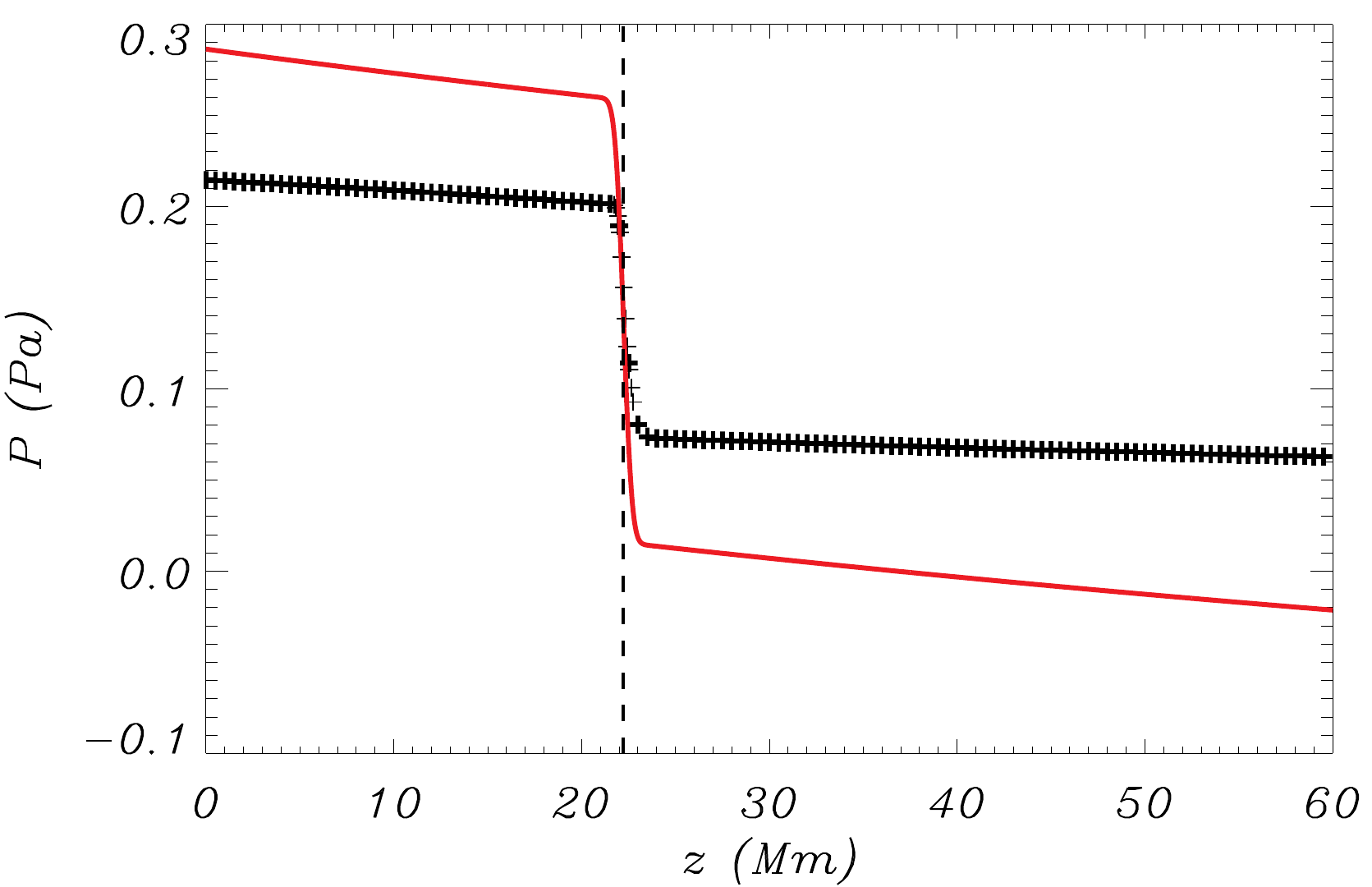}}
		\caption{Pressure as a function of height. Black symbols represent the numerical results from simulations with $B_{0} = 5 \ \rm{G}$. Red lines show the analytical results given by Eq. (\ref{eq:pc}). Top: case with $\rho_{\rm{b0}} = 10^{-10} \ \rm{kg \ m^{-3}}$. Bottom: case with $\rho_{\rm{b0}} = 10^{-9} \ \rm{kg \ m^{-3}}$. The vertical black dashed lines mark the positions of the condensations.}
		\label{fig:pgrad}
	\end{figure}

	In Fig. \ref{fig:pgrad}, we compare the analytical profile derived above with vertical cuts of the pressure at $x = 0$ from two different simulations in which the magnetic field strength is $B_{0} = 5 \ \rm{G}$. Both panels show the results at a time $t = 500 \ \rm{s}$, but the top panel corresponds to a simulation with $\rho_{\rm{b0}} = 10^{-10} \ \rm{kg \ m^{-3}}$, while in the bottom panel the density is 10 times larger, i.e., $\rho_{\rm{b0}}= 10^{-9} \ \rm{kg \ m^{-3}}$. The black symbols represent the data from the simulations and the red lines represent the analytical results. The dashed vertical lines mark the position of the condensation.
	
	For the case of smaller density, the numerical and the analytical results overlap, meaning that the condition for the constant-speed phase is fulfilled. The magnitude of the pressure gradient is enough to balance the action of gravity. On the contrary, the bottom panel shows that a larger density requires a larger pressure gradient to reach the constant-speed phase. The pressure gradient computed from the simulation is larger than in the previous case but it is still not enough to completely compensate the acceleration of gravity. Therefore, the blob is still gaining speed at this time of the simulation. Consequently, it has reached a lower height than the lighter condensation of the top panel.
	
\bibliographystyle{aa}
\bibliography{cr_bib}

\end{document}